\documentclass[final,3p,times,authoryear]{elsarticle}

\usepackage{amsmath,amssymb}
\usepackage{graphicx}
\usepackage{placeins}
\usepackage{hyperref}
\usepackage{float}
\usepackage{ragged2e}
\usepackage[figuresright]{rotating}

\journal{New Astronomy}

\begin{document}

\begin{frontmatter}

\title{A Multiband Catalog of Infrared-Excess Sources in the Vela--Carina Region: VVVX, DECaPS, and GLIMPSE}

\author[aff1,aff2]{R. M. Torres}
\ead{roberto.torres@ifc.edu.br}

\author[aff1]{R. K. Saito}
\author[aff1,aff6]{V. Fermiano}
\author[aff1]{P. Esteves}
\author[aff1,aff3]{B. W. Borges}
\author[aff1]{D. Quispe}
\author[aff4,aff5]{D. Minniti}

\affiliation[aff1]{organization={Departamento de Física, Universidade Federal de Santa Catarina},
  addressline={Trindade},
  city={Florianópolis},
  postcode={88040-900},
  state={Santa Catarina},
  country={Brazil}}

\affiliation[aff2]{organization={Instituto Federal Catarinense},
  postcode={89703-720},
  city={Concórdia},
  state={Santa Catarina},
  country={Brazil}}

\affiliation[aff3]{organization={Coordenadoria Especial de Física, Química e Matemática, Universidade Federal de Santa Catarina},
  addressline={Jardim das Avenidas},
  city={Araranguá},
  state={Santa Catarina},
  country={Brazil}}

\affiliation[aff4]{organization={Instituto de Astrofísica, Dep. de Física y Astronomia, Facultad de Ciencias Exactas, Universidad Andres Bello},
  addressline={Av. Fernández Concha 700},
  city={Santiago},
  country={Chile}}

\affiliation[aff5]{organization={Vatican Observatory},
  addressline={Specola Vaticana, V-00120},
  city={Vatican City},
  country={Vatican City State}}

\affiliation[aff6]{
  organization={Universidad de Valparaíso},
  city={Valparaíso},
  country={Chile}
}

\begin{abstract}
Large multiwavelength surveys of the Galactic midplane enable the identification of sources exhibiting infrared excess, a signature of circumstellar material; however, this task is hindered by high extinction and severe source crowding.

We aim to construct a homogeneous catalog of infrared-excess candidates in a $\sim$7.0 deg$^{2}$ field of the Vela--Carina region and to characterize the nature of their excess emission through spectral energy distribution (SED) analysis.

We combine optical photometry from the Dark Energy Camera Plane Survey (DECaPS), near-infrared (NIR) photometry from the VISTA Variables in the Vía Láctea eXtended survey (VVVX), and mid-infrared (MIR) photometry from the Spitzer Galactic Legacy Infrared Mid-Plane Survey Extraordinaire (GLIMPSE) using hierarchical positional cross-matching. Infrared-excess candidates are selected based on photometric quality criteria and MIR color diagnostics. Their SEDs are analyzed with the Virtual Observatory SED Analyzer (VOSA) using grids of theoretical stellar atmosphere models. Photogeometric distance estimates were adopted for the physical characterization of the sources, while variability information from Gaia Data Release 3 (Gaia DR3) and the Near-Earth Object Wide-field Infrared Survey Explorer Reactivation mission (NEOWISE) was used as a complementary diagnostic.

From the matched VVVX--GLIMPSE--DECaPS catalog of 249\,719 sources, we identify 1\,512 objects exhibiting MIR excess, of which 191 have complete 12-band photometry with $\mathrm{S/N}>10$, enabling detailed SED analysis. For sources with available estimates, the adopted photogeometric distances are statistically consistent with the corresponding geometric distances for nearly the entire sample.

The SED analysis independently supports the MIR-excess nature of the majority of the selected sources, indicating infrared emission beyond that expected from stellar photospheres alone. Approximately 15\% of the sources are flagged as variable by \textit{Gaia}, while NEOWISE light curves reveal MIR variability at the level of a few tenths of a magnitude. Comparisons with previously published young stellar object (YSO) and variability-selected catalogs show varying degrees of overlap, with particularly limited overlap with variability-selected samples, indicating that MIR-excess selection provides a complementary route to identifying candidate YSOs in the Vela--Carina region.

The resulting catalog provides a well-defined sample of MIR-excess sources with homogeneous multiband photometry, SED characterization, distance information, and variability diagnostics, and constitutes a robust foundation for future spectroscopic and time-domain studies.
\end{abstract}

\begin{keyword}
infrared-excess sources \sep star formation \sep near-infrared photometry \sep multiband surveys \sep Vela--Carina region \sep astronomical catalogs
\end{keyword}

\end{frontmatter}

% SECTION 1 ---------------------------------------------------------------------------
\section{Introduction}
% ---------------------------------------------------------------------------

Near-infrared (NIR) and mid-infrared (MIR) observations are particularly effective for characterizing stellar populations in heavily obscured star-forming regions, where extinction and crowding limit optical diagnostics \citep{LadaWilking1984}. Emission in excess of a purely photospheric spectrum at wavelengths longer than $\sim 2\,\mu$m is commonly interpreted as evidence of circumstellar dust and is routinely associated with pre-main-sequence (PMS) objects spanning protostellar to disk-bearing stages \citep{Adams1987,Evans2009,Dunham2015}. Building statistically meaningful samples of infrared-excess sources therefore provides a direct route to mapping recent star formation and locating embedded populations across the Milky Way.

The VISTA Variables in the V\'ia L\'actea survey \citep[VVV;][]{Minniti2010} and its extension VVVX \citep{Saito2024} provide deep $JHK_s$ photometry across the southern bulge and disk, with multi-epoch coverage and a time baseline exceeding a decade. The expanded VVVX footprint toward the outer disk and lower latitudes increases the parameter space for identifying obscured objects and for linking infrared colors with time-domain behavior.

To maximize the reliability of infrared-excess identification in highly reddened inner-disk fields, NIR photometry alone is insufficient, as reddened field stars can mimic excess colors in limited band combinations. Incorporating optical constraints from DECaPS ($grizY$; \citealt{Saydjari2023}) and MIR measurements from \textit{Spitzer}/GLIMPSE (3.6--8.0\,$\mu$m; \citealt{Benjamin2003}) enables multi-color selection that is substantially less degenerate with extinction and provides the photometric leverage required for basic SED consistency checks.

Several studies have investigated young stellar populations and infrared-excess sources in the Vela--Carina region using different observational strategies. \citet{Povich2011} identified candidate YSOs in the Carina Nebula primarily through MIR excess emission detected with \textit{Spitzer}. Using VISTA and \textit{Spitzer} photometry, \citet{Zeidler2016} identified infrared-excess sources in part of the Vela--Carina region through color-based selection criteria. \citet{Damiani2017} characterized the stellar populations of the Trumpler~14 and Trumpler~16 clusters using Gaia--ESO spectroscopic observations combined with complementary photometric data. On a larger Galactic scale, the Spitzer/IRAC Candidate YSO Catalog for the Inner Galactic Midplane (SPICY; \citealt{Kuhn2021}) provides a homogeneous compilation of candidate YSOs identified from \textit{Spitzer}/IRAC observations, while \citet{Borissova2025} identified YSO candidates in the VVVX survey using NIR variability complemented by spectroscopic information. Together, these studies illustrate the complementary nature of infrared-excess, spectroscopic, and variability-based approaches for identifying young stellar populations.

In the context of these previous studies, we compile a homogeneous 12-band catalog by combining VVVX, DECaPS, and GLIMPSE photometry over a $\sim$7\,deg$^2$ region of the southern Galactic midplane centered near $(\ell,b)=(285.8^\circ,-0.5^\circ)$. This field is characterized by high source density and strong, spatially variable extinction, and lies in the general direction of the Vela--Carina star-forming complex. Our primary objective is to identify a robust sample of MIR-excess candidates through conservative photometric filtering, hierarchical cross-matching, and empirical color--color diagnostics. By adopting a homogeneous multiwavelength selection procedure, the resulting catalog complements previous studies based on infrared-excess, spectroscopic, or variability selection criteria and provides a foundation for subsequent investigations of variability, clustering, and the physical nature of the selected sources.

This study is complementary to the recent work of \citet{Esteves2026}, which investigates infrared-excess sources identified in the VVV survey through SED characterization and variability analysis.

% SECTION 2 +++++++++++++++++++++++++++++++++++++++++++
\section{Data and survey cross-matching}
% +++++++++++++++++++++++++++++++++++++++++++

We assembled optical-to-MIR photometry for a southern Galactic midplane field by combining three large-area surveys. NIR $JHK_s$ measurements are taken from VVVX, reaching $K_s \sim 18$\,mag with a typical image quality of $\sim 0.8^{\prime\prime}$ \citep{Saito2024}. Optical $grizY$ photometry is adopted from DECaPS \citep{Saydjari2023}, which is optimized for crowded Galactic-plane regions. MIR data at 3.6, 4.5, 5.8, and 8.0\,$\mu$m are taken from the GLIMPSE survey obtained with Spitzer/IRAC \citep{Benjamin2003}.

The analyzed area covers approximately 7~deg$^2$, centered at $(\ell, b) = (285.8^\circ,-0.5^\circ)$ and spanning $282.0^\circ \le \ell \le 289.0^\circ$ and $-1.0^\circ \le b \le 0.0^\circ$. These sky coordinates fall within five adjacent VVVX tiles (d1082--d1086), part of the Disk230 extension of the survey \citep{Saito2024}. Although these tiles together cover 8.23~deg$^2$ in their nominal footprint, our analysis is restricted to the 7.0~deg$^2$ region where VVVX, DECaPS, and GLIMPSE overlap completely, ensuring uniform multi-band coverage.

% SUBSECTION 2.1 ---------------------------------------------------------------------------
\subsection{Matched multi-survey catalog}
% ---------------------------------------------------------------------------
We used VVVX as the positional reference frame and performed cross-matching in two stages. First, VVVX sources were associated with GLIMPSE counterparts using a $1^{\prime\prime}$ search radius; a second pass with a $5^{\prime\prime}$ radius was then used to inspect and resolve cases with ambiguous or multiple matches. The resulting VVVX--GLIMPSE table was subsequently matched to DECaPS using the same two-step strategy. When more than one candidate counterpart was available, we retained the entry with the smallest quoted photometric uncertainty. The final merged catalog therefore preserves a single, internally consistent 12-band photometric measurement per source.

The GLIMPSE source identifiers reported throughout this work follow the official nomenclature of the Infrared Science Archive (IRSA) \textit{Vela--Carina Archive} (SSTVELA designation), from which the GLIMPSE data were retrieved.

The main sample-selection and analysis steps, together with the corresponding sample sizes, are summarized in Table~\ref{tab:sample_summary}.

\begin{table}[ht]
\centering
\caption{Summary of the main sample-selection and analysis steps adopted in this work. S/N denotes signal-to-noise ratio.}
\label{tab:sample_summary}
\begin{tabular}{lr}
\hline
Selection or analysis step & Number of sources \\
\hline
VVVX sources in the analyzed field & $\sim$3.75 million \\
Matched VVVX--GLIMPSE--DECaPS catalog & 249\,719 \\
S/N $>10$ in $J$, $H$, $K_s$, and [4.5] & 27\,610 \\
MIR-excess candidates & 1\,512 \\
MIR-excess sources with S/N $>10$ in all 12 bands & 191 \\
Sources with distance and extinction estimates & 181 \\
Sources with MIR light curves & 172 \\
\hline
\end{tabular}
\end{table}

\begin{figure*}[ht!]
  \centering
  \includegraphics[width=0.98\textwidth]{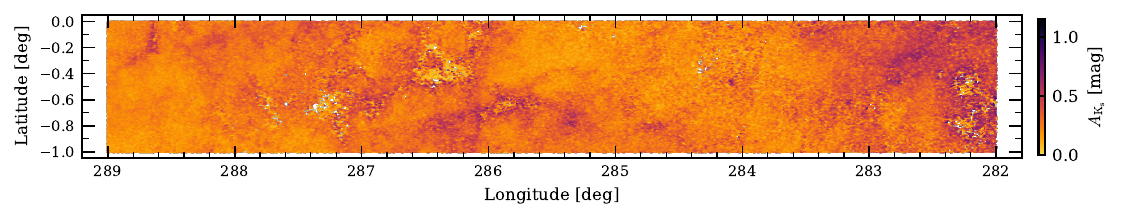}
  \caption{Spatial map of the NIR extinction $A_{K_s}$ across the analyzed VVVX field, constructed from 249\,719 sources matched between the VVVX, GLIMPSE, and DECaPS surveys.  For each source, the visual extinction $A_V$ was converted to the extinction in the $K_s$ band ($A_{K_s}$) using the extinction coefficients of \citet{Catelan2011}, based on the extinction law of \citealt{Cardelli1989}, adopting $A_{K_s}/A_V = 0.118$. The sources were projected onto the Galactic $(\ell,b)$ plane and mapped onto a two-dimensional pixel grid. Each source contributes to a square pixel kernel of $5\times5$ pixels centered on its projected position, and the value displayed in each pixel corresponds to the mean $A_{K_s}$ of the sources contributing to that pixel.}
  \label{fig:AKs_map}
\end{figure*}

A first visual impression of the spatially varying extinction across the VVVX field is provided in Fig.~\ref{fig:AKs_map}. The detailed procedure used to compute this map is described in Section~\ref{sec:extinction}.

% -------------------------------------------------------------------------- FIGURE 1
% -------------------------------------------------------------------------- end FIGURE 1

% SUBSECTION 2.2 ---------------------------------------------------------------------------
\subsection{Line-of-sight extinction estimation}
% ---------------------------------------------------------------------------
\label{sec:extinction}

For each matched source, photogeometric distances were taken from the catalog of \citet{BailerJones2021} (hereafter BJ21), which is based on Gaia Early Data Release 3 (Gaia EDR3) sources. The selected objects were subsequently associated with Gaia Data Release 3 (Gaia DR3) using the Gaia source identifier in order to retrieve astrometric and variability-related parameters. The line-of-sight visual extinction $A_V$ was then retrieved from the three-dimensional dust map of \citet{Zucker2025} (hereafter Z25), evaluated at the corresponding BJ21 distance. These astrometric quantities were incorporated only after the photometric selection and were not used to define the 12-band catalog. The use of distance-dependent extinction is fundamental in this region of the Galactic midplane, where the dust distribution varies strongly along the line of sight.

To characterize the global behavior of the sample prior to constructing the extinction map, we examined the distributions of distances and visual extinctions for all matched sources. The bin widths for both histograms were computed using the Freedman--Diaconis rule \citep{FreedmanDiaconis1981}, which provides an optimal balance between resolution and robustness to outliers. Only finite entries were retained, and the \emph{mode}, defined as the center of the most populated histogram bin, was adopted as the representative value of each distribution.

% -------------------------------------------------------------------------- FIGURE 2
\begin{figure*}[ht!]
    \centering
    \includegraphics[width=0.75\columnwidth]{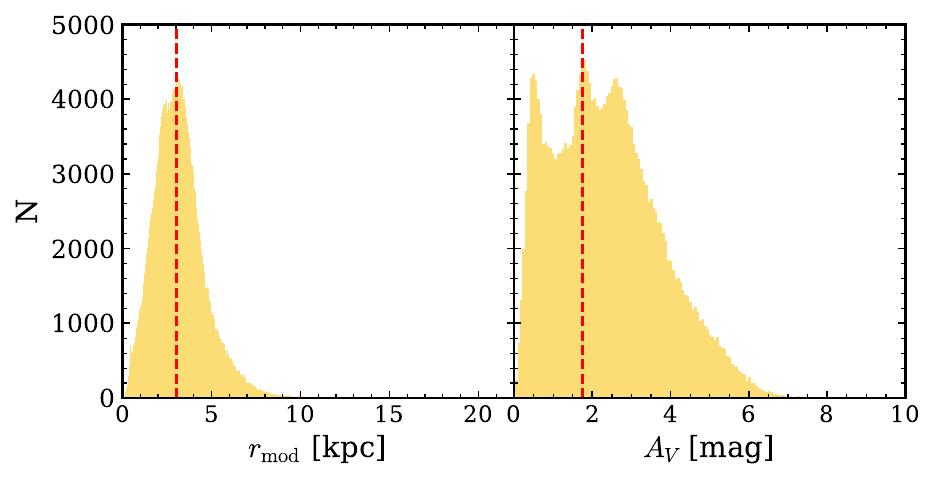}
    \caption{Distributions of photogeometric distances (left) from BJ21 and visual extinctions (right) from the 3D dust map of Z25 for the 249\,719 matched sources. A modal estimator is used to characterize each distribution, and the red dashed lines indicate the corresponding modal values.}
    \label{fig:distance_extinction_histograms}
\end{figure*}
% -------------------------------------------------------------------------- end FIGURE 2

As shown in Fig.~\ref{fig:distance_extinction_histograms}, the catalog exhibits modal values of $r_{\rm mod}=3048$\,pc and $A_{V,{\rm mod}}=1.751$\,mag. These characteristic values are later used to shift the PAdova and TRieste Stellar Evolution Code (PARSEC) isochrones and to compute the reddening vectors employed in the color-magnitude diagram (CMD) and color-color diagram (CCD) analyses.

The conversion from $A_V$ to NIR extinction in the VISTA bands was performed using the coefficients listed in Table~\ref{tab:extinction_coeff}, adopted from \citet{Catelan2011} and based on the extinction law of \citet{Cardelli1989}. In particular, the coefficient $A_{K_s}/A_V$ was used to derive the extinction in the $K_s$ band for each source.

\begin{table}[ht]
\caption{Adopted extinction coefficients $A_X/A_V$ for the VISTA $J$, $H$, and $K_s$ bands, taken from \citet{Catelan2011} and based on the extinction law of \citet{Cardelli1989}.}
\label{tab:extinction_coeff}
\centering
\begin{tabular}{lcc}
\hline\hline
Filter & $\lambda_{\rm eff}$ ($\mu$m) & $A_X/A_V$ \\
\hline
$J$     & 1.254 & 0.280 \\
$H$     & 1.646 & 0.184 \\
$K_s$   & 2.149 & 0.118 \\
\hline
\end{tabular}
\end{table}

Individual $A_{K_s}$ values were projected onto the Galactic $(\ell,b)$ plane and combined on a two-dimensional pixel grid to construct the extinction map shown in Fig.~\ref{fig:AKs_map}. Each source contributes to a small square pixel kernel around its projected position, and the value displayed in each pixel corresponds to the mean $A_{K_s}$ of the contributing sources. This representation reveals the strong spatial variations in extinction across the VVVX field and provides the extinction estimates required for the CMD analysis and for the subsequent construction of SEDs of the MIR-excess sources.

% SECTION 3 ---------------------------------------------------------------------------
\section{MIR-excess selection}
% ---------------------------------------------------------------------------
\label{sec:mir_selection}

To identify MIR-excess sources, we constructed the $(J-H)$ versus $(K_s-[4.5])$ color--color diagram using the merged VVVX--GLIMPSE--DECaPS catalog. We restricted the analysis to measurements with signal-to-noise ratio $(\mathrm{S/N}) > 10$ in $J$, $H$, $K_s$, and $[4.5]$, yielding a parent sample of 27\,610 objects with valid photometric measurements in the four relevant bands. This high-quality subset defines a narrow, extinction-dominated stellar locus that can be characterized empirically.

We quantified the locus by binning the sample in $(J-H)$ and modeling the corresponding $(K_s-[4.5])$ distributions with Gaussian profiles. The peak value in each bin traces the dominant field population; these modal points were then fitted with a linear relation, $(J-H)= a_{\rm fit} + b_{\rm fit}\,(K_s-[4.5])$. For the present field, we obtain $a_{\rm fit}=0.3145 \pm 0.0032$ and $b_{\rm fit}=2.1540 \pm 0.0078$. This empirical sequence is used as a reference against which MIR-excess deviations are evaluated.

MIR-excess candidates were identified following the geometric criterion introduced by \citet{Zeidler2016}. A source is classified as exhibiting MIR excess when it simultaneously satisfies:
\begin{align}
(K_s - [4.5]) &> 0.49, \\
(J - H)       &\ge 0.70, \\
(J - H)       &< b_{\rm fit}\,\bigl((K_s - [4.5]) - 0.49\bigr) + 0.70.
\end{align}

% -------------------------------------------------------------------------- FIGURE 3
\begin{figure*}[ht!]
  \centering
  \includegraphics[width=0.85\columnwidth]{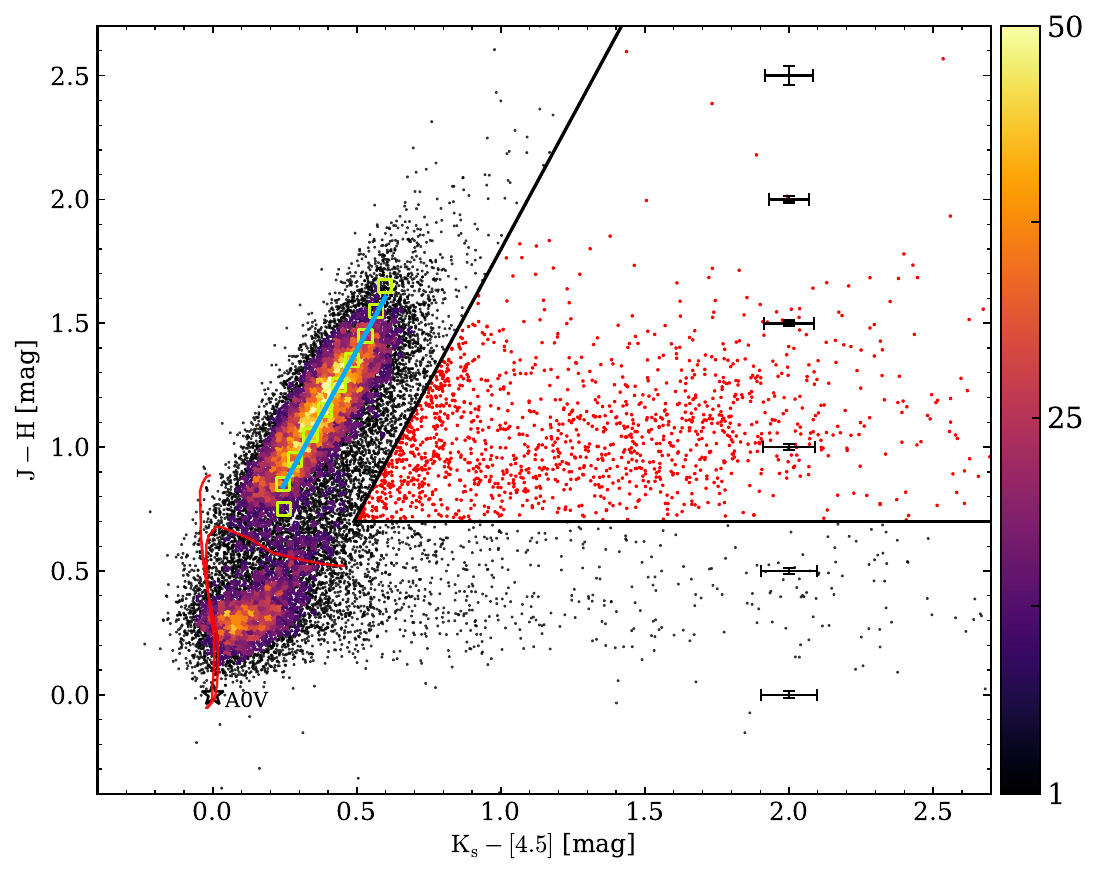}
  \caption{Color--color diagram $(J-H)$ versus $(K_s-[4.5])$ for the 27\,610 sources with $\mathrm{S/N}>10$ in the four relevant bands. The background density is shown as a Hess diagram, while MIR-excess candidates identified through the Zeidler criterion are plotted in red. The blue line denotes the empirical reddening locus derived from modal slice fitting, and yellow squares mark the corresponding modal points. Representative photometric uncertainties are shown on the right-hand side. A PARSEC v1.2S isochrone (solar metallicity, 316\,Myr) is overplotted as a morphological reference.}
  \label{fig:ccd_mir}
\end{figure*}
% -------------------------------------------------------------------------- end FIGURE 3

The first two conditions restrict the selection to redder objects, while the third identifies those lying significantly to the right of the reddening locus, i.e., sources exhibiting an excess at 4.5\,$\mu$m that cannot be explained by interstellar reddening alone. Applying this criterion yields 1\,512 MIR-excess objects, whereas the remaining 26\,098 form the background population used to map the CCD density structure.

Sources located close to the adopted MIR color-selection boundaries, particularly those involving the [4.5]~$\mu$m band, are more sensitive to photometric uncertainties. Throughout the analysis, special attention is therefore given to objects lying near the edges of the selection criteria.

Given the high source density in the Galactic plane, especially in the Vela region, potential effects of crowding and spurious associations in the \textit{Spitzer}/IRAC data cannot be fully excluded. These effects are expected to be more relevant for marginal cases and are taken into account when interpreting the infrared excess properties.

Figure~\ref{fig:ccd_mir} presents the final CCD, combining the Hess diagram of the field population, the MIR-excess sources, the empirical reddening sequence, the modal slice points, and representative photometric uncertainties. For visual guidance, a PARSEC v1.2S isochrone (solar metallicity, Two Micron All Sky Survey (2MASS) + Infrared Array Camera (IRAC) photometric system, age $\sim316$ Myr) is overplotted to indicate the approximate locus of main-sequence stars. This integrated framework provides a robust, observation-based basis for identifying MIR-excess sources across the VVVX field.

% SECTION 4 ---------------------------------------------------------------------------
\section{Global photometric context from the VVVX survey}
% ---------------------------------------------------------------------------

\begin{figure*}[ht!]
\centering
\includegraphics[width=0.95\columnwidth]{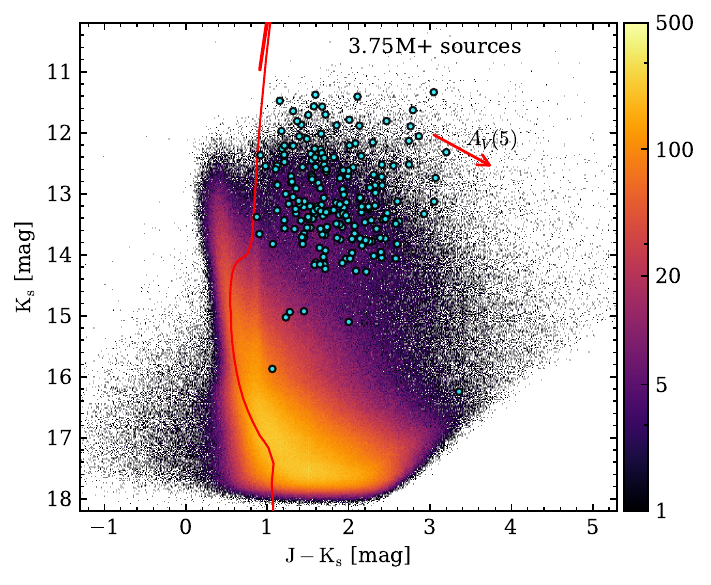}
\caption{Color--magnitude diagram (CMD) of the VVVX field used in this work. The background density map shows the full set of $3.75$ million VVVX sources with valid $J$ and $K_s$ photometry, displayed as a logarithmic Hess diagram. Overplotted in cyan are the 191 MIR-excess sources constituting the final sample, all of which have signal-to-noise ratios $\mathrm{S/N} > 10$ in each of the 12 photometric bands used in the SED analysis. A PARSEC v1.2S isochrone (solar metallicity, shifted to $d = 3048$\,pc and $A_V = 1.751$\,mag) is shown in red as a morphological reference. The red arrow indicates the extinction vector corresponding to $A_V = 5$\,mag, computed using the VISTA extinction law of \citet{Catelan2011}.}
\label{fig:cmd_vvvx_plus_191}
\end{figure*}

To characterize the stellar population of the field and provide a reference framework for the identification of MIR-excess sources, we constructed the color--magnitude diagram (CMD) $(J-K_s)$ versus $K_s$ shown in Fig.~\ref{fig:cmd_vvvx_plus_191}. The background density map represents the full set of 3.75 million VVVX sources with reliable $J$ and $K_s$ photometry, displayed as a logarithmic Hess diagram. This large sample traces the underlying Galactic distribution of dwarfs and giants, together with the strong effects of interstellar extinction along the line of sight. The CMD exhibits the characteristic morphology expected for a highly reddened inner-disk field, including the foreground main sequence, the broadened red clump, and the extended giant branch.

Shown in cyan are the 191 MIR-excess sources constituting the final sample, selected from the 1\,512 candidates identified in Sect.~3. All of these sources have signal-to-noise ratios $\mathrm{S/N}>10$ in each of the twelve photometric bands (DECaPS $grizY$, VVVX $JHK_s$, and \textit{Spitzer}/IRAC 3.6, 4.5, 5.8, and 8.0~$\mu$m) used to construct the SEDs. Their distribution in the CMD shows that they form a relatively coherent population, primarily located at $K_s \simeq 12$--15.5\,mag and $(J-K_s) \simeq 1$--2.5\,mag, with only a few objects extending to $(J-K_s)\sim 3$\,mag. This locus is broadly consistent with moderately reddened stellar populations, although the physical nature of the sources may be diverse. The relatively compact clustering of the MIR-excess candidates in the CMD supports the robustness of the photometric selection.

Several previous studies have identified YSOs and infrared-excess sources in the Vela--Carina region using different observational strategies and selection criteria. To place our catalog in the context of these surveys, we performed positional cross-matches with the principal published catalogs using a matching radius of $1^{\prime\prime}$.

The comparison with the MIR-excess catalog of \citet{Zeidler2016} yielded 31 positional counterparts. The Galactic-coordinate range covered by their published catalog overlaps only part of the region analyzed in this work, encompassing 56 of the 191 MIR-excess sources identified here. Within this common sky area, 31 positional counterparts were identified. We also compared our catalog with the SPICY catalog of \citet{Kuhn2021}, obtaining 125 positional counterparts. Since the SPICY catalog encompasses the full Galactic-coordinate range of our catalog, all 191 MIR-excess sources lie within its spatial coverage. The remaining unmatched sources can be attributed, at least in part, to differences in source selection, photometric constraints, and classification criteria.

For completeness, we also compared our catalog with the Pan-Carina YSO catalog of \citet{Povich2011}, obtaining two positional counterparts. Only ten of our sources are located within the Galactic-coordinate range occupied by the published Povich catalog. In contrast, no positional counterparts were found in the stellar sample of \citet{Damiani2017}. This result is not unexpected because the Damiani catalog was constructed using membership and spectroscopic diagnostics rather than homogeneous MIR-excess selection. Finally, a comparison with the VVVX variability-selected YSO catalog of \citet{Borissova2025} yielded a single common object (source \#170 in our catalog), illustrating the complementary nature of infrared-excess and variability-based approaches for identifying young stellar populations.

Astrometric and auxiliary parameters from \textit{Gaia} Data Release 3 (Gaia DR3), together with photogeometric distances from BJ21 and line-of-sight extinction estimates from Z25, were incorporated only after the photometric selection. These datasets were not included in the cross-match defining the 12-band catalog, but are used exclusively to refine distances, proper motions, and extinction estimates for the final MIR-excess sample.

Among the 191 MIR-excess sources, 167 (87\%) have $\mathrm{RUWE}<1.4$, whereas 24 (13\%) have $\mathrm{RUWE}\geq1.4$. Forty-one sources have formally negative Gaia parallaxes. However, 35 have $\varpi/\sigma_{\varpi}>-2$, indicating that their negative parallaxes are consistent with the measurement uncertainties and are not significant at the 2$\sigma$ level. Photogeometric distances from BJ21 are available for 40 of these 41 sources. The remaining object (Source~163) has no corresponding BJ21 distance estimate; consequently, no distance-dependent extinction from Z25 is assigned to this source. In the complete sample, valid BJ21 photogeometric distances are available for 181 of the 191 sources. For the remaining ten objects, no distance-dependent extinction estimate from Z25 is available. Consequently, VOSA-derived physical parameters, photogeometric distances, and Z25 extinction values are reported only for the 181 sources with valid BJ21 distance estimates.

For the 181 sources with both geometric and photogeometric distances from BJ21, the mean ratio $r_{\rm photogeo}/r_{\rm geo}$ is 1.08 (corresponding to a mean logarithmic difference of 0.022 dex). Their respective 68\% confidence intervals overlap for 179 sources (98.9\%), indicating that the two distance estimates are statistically consistent for almost the entire sample. Accordingly, the photogeometric distances from BJ21 were adopted throughout this work, as they provide more reliable distance estimates than direct geometric distances for sources with low-significance or negative \textit{Gaia} parallaxes.

For reference, a PARSEC v1.2S isochrone for solar metallicity, shifted to a distance of 3048\,pc and an extinction of $A_V = 1.751$\,mag, is shown in red. Although the isochrone is not used in the MIR-excess identification itself, it provides a morphological guide to the intrinsic stellar locus expected for a typical stellar population at the inferred distance.

The isochrone is used only as a morphological reference and does not imply a specific evolutionary stage for the sources exhibiting MIR-excess.

% -------------------------------------------------------------------------- FIGURE 4
% -------------------------------------------------------------------------- end FIGURE 4

The red arrow in Fig.~\ref{fig:cmd_vvvx_plus_191} illustrates the extinction vector corresponding to $A_V = 5$\,mag. The direction and magnitude of this vector highlight the strong influence of interstellar extinction on the CMD morphology.

Overall, Fig.~\ref{fig:cmd_vvvx_plus_191} establishes the global photometric context within which the MIR-excess candidates are identified. The combination of the full VVVX density distribution, the final 191-source sample, and the evolutionary and extinction references provides a clear visual framework for interpreting the subsequent analysis.

% SECTION 5 ---------------------------------------------------------------------------
\section{SED analysis with VOSA}
% ---------------------------------------------------------------------------
\label{sec:vosa}

The SEDs of the 191 sources selected through the MIR color criteria were analyzed using the Virtual Observatory  SED Analyzer (VOSA; \citealt{Bayo2008}). Although the SEDs of all 191 sources were inspected, physical parameters from the VOSA fitting are reported only for the 181 sources with valid BJ21 photogeometric distances and corresponding distance-dependent Z25 extinction estimates. The primary objective of this analysis is to characterize the stellar photospheric emission and to independently assess, on a source-by-source basis, the presence and relative strength of infrared excess emission, complementing the photometric selection described in Sect.~\ref{sec:mir_selection}. 

Although the infrared excess properties are discussed here in a global sense, individual SEDs were also examined to assess the robustness of the excess emission in each source. This approach is particularly important for objects with marginal infrared colors, allowing us to evaluate the consistency between color--color diagnostics and the overall SED morphology.

The SED fitting was performed using several stellar atmosphere model grids available within VOSA. As the primary reference grid, we adopted the BT-Settl models \citep{Allard2012}, which provide extensive coverage in effective temperature and surface gravity suitable for PMS stars across the full parameter space explored in this work. For consistency checks, the BT-Settl CIFIST grid \citep{Allard2012,Caffau2011} was also employed for sources with effective temperatures up to $\sim7000$~K, allowing an assessment of the impact of different solar abundance prescriptions on the derived stellar parameters.

% -------------------------------------------------------------------------- FIGURE 5
\begin{figure*}[ht!]
\centering
\includegraphics[width=0.98\columnwidth]{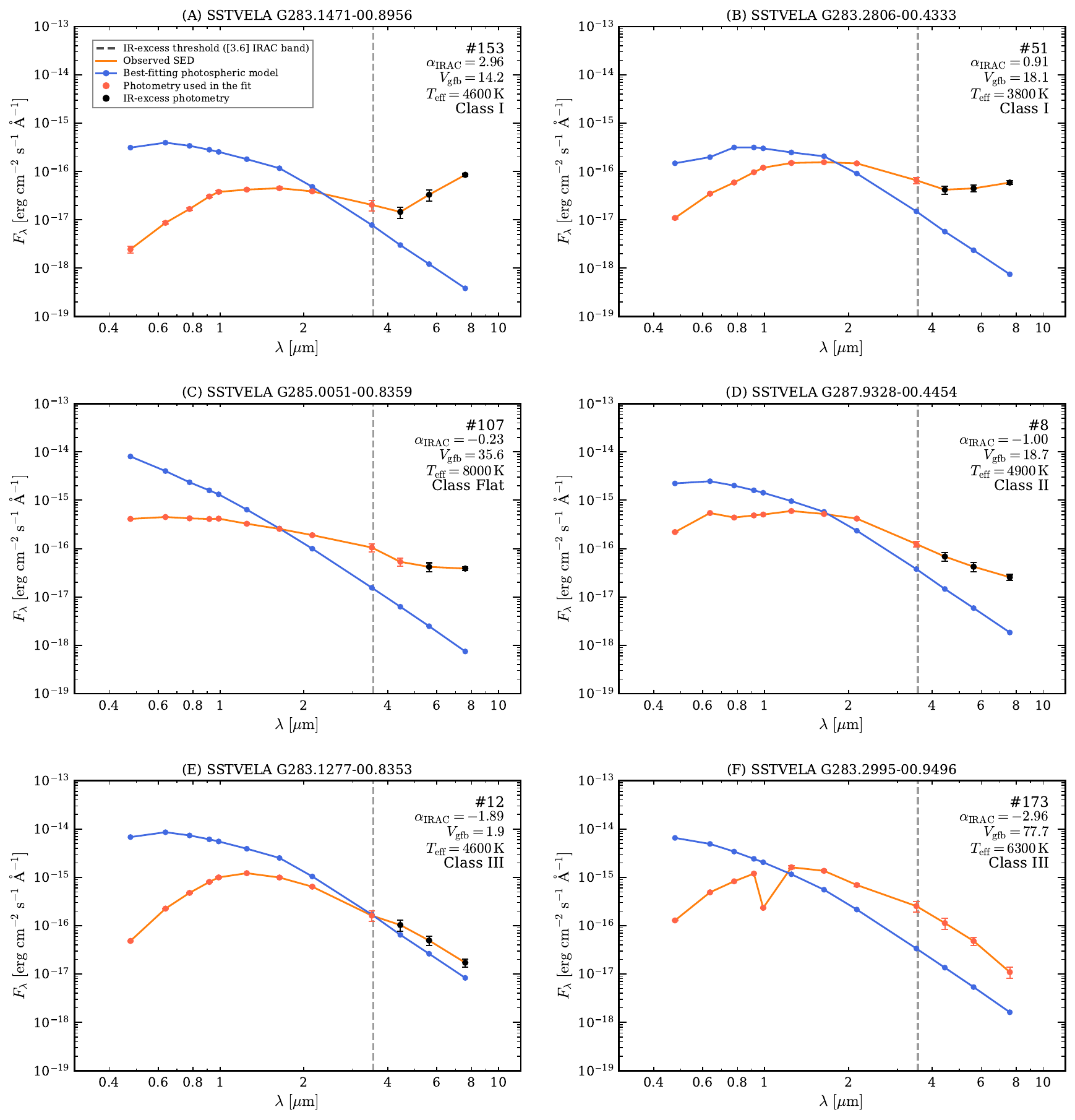}
\caption{Representative SEDs of six sources selected from the final sample to illustrate the range of infrared properties identified in the VOSA analysis. Solid blue lines correspond to the best-fitting stellar atmosphere models, while solid orange lines represent the observed SEDs interpolated between the photometric measurements. Red points indicate the photometric data used in the fit, whereas black points indicate photometric measurements identified as infrared excess according to the adopted excess criterion. The vertical error bars represent three times the corresponding photometric uncertainties ($3\sigma$), following the convention adopted in the VOSA graphical output. The vertical dashed line marks the adopted wavelength threshold above which the infrared-excess search was performed. The panels are arranged in decreasing order of the IRAC spectral index, from the highest to the lowest $\alpha_{\mathrm{IRAC}}$ values, illustrating progressively weaker levels of MIR excess. The labels ``Class~I'', ``Flat'', ``Class~II'', and ``Class~III'' refer to the IRAC spectral-index classification based on $\alpha_{\mathrm{IRAC}}$ (see Sect.~6).}
\label{fig:sed_mosaic_1}
\end{figure*}
% -------------------------------------------------------------------------- end FIGURE 5

In addition, Kurucz ATLAS9 atmospheric models \citep{CastelliKurucz2003} (ODFNEW/NOVER) were included in the fitting procedure, primarily to account for hotter PMS candidates and intermediate-mass objects. Although these models are not intended to represent the full PMS regime, they provide an adequate description of the stellar photosphere at higher effective temperatures and serve as a complementary grid for consistency checks. Within the uncertainties inherent to broadband SED fitting, the different atmospheric grids yield consistent results in terms of photospheric shape and infrared excess identification. While the choice of model can influence the detailed values of individual stellar parameters, it does not affect the detection of significant MIR excess, which is the primary goal of this analysis. The explored parameter ranges were $T_{\rm eff}=2500$--12000~K, $\log g=3.0$--4.5, and $[{\rm M/H}]=-0.5$ to $+0.3$ for the BT-Settl models; $T_{\rm eff}=2500$--7000~K, $\log g=3.0$--4.5, with solar metallicity, for the BT-Settl CIFIST models; and $T_{\rm eff}=4500$--9000~K, $\log g=3.5$--5.0, $[{\rm M/H}]=-0.5$ to $+0.2$, with $\alpha=0.4$, for the Kurucz ODFNEW/NOVER models.

The fitting procedure was carried out primarily without imposing external constraints on distance or extinction. This approach allows a direct comparison between the observed photometry and purely photospheric models and provides a robust diagnostic of MIR excess emission, independent of assumptions on line-of-sight extinction or distance. In a secondary step, the fitting was repeated with distance and extinction fixed in order to verify the consistency of the results. 

The quality of the SED fits was evaluated using the modified goodness-of-fit parameter $V_{\mathrm{gfb}}$, which reduces the impact of underestimated photometric uncertainties. Sources exhibiting MIR excess systematically show higher $V_{\mathrm{gfb}}$ values, reflecting the inability of purely photospheric models to reproduce the observed emission at longer wavelengths. Although fixing distance and extinction generally reduces the absolute values of $V_{\mathrm{gfb}}$, the relative ranking of the sources is preserved, confirming the robustness of the excess identification.

Using the BT-Settl models, the stellar photospheric emission is well reproduced from the optical to the NIR for most sources, while a clear excess is detected at MIR wavelengths for the majority of the sample. This excess cannot be explained by photospheric emission alone and is interpreted as arising from warm circumstellar material. Although the SED fits favor cool stellar photospheres for a large fraction of the sample, the fitting procedure was allowed to explore higher effective temperatures to accommodate intermediate-mass PMS objects, such as Herbig~Ae/Be stars. A PMS interpretation for these hotter candidates is supported by a combination of surface gravity estimates and the presence of MIR excess, both inconsistent with diskless main-sequence stars.

While all sources satisfy the MIR color-selection criteria, the SED analysis reveals a significant diversity in the strength of the infrared excess. In particular, a small subset of objects is well reproduced by stellar photospheric models up to the IRAC wavelengths, showing only weak or marginal excesses within the uncertainties of the available photometry. These sources are retained in the catalog for completeness, but the interpretation of their infrared properties should consider both the overall SED morphology and the IRAC spectral index.

The best-fit results obtained with VOSA were retrieved in VOTable format and adopted as the reference catalog for this work. The stellar parameters derived from the SED analysis, including effective temperature, surface gravity, bolometric luminosity, and the atmospheric model providing the best fit, are reported in Appendix~A (Table~A.1). The complete catalog, including the full set of photometric measurements, astrometric information, extinction estimates, and SED fitting results, is available at the CDS.

% SECTION 6 ---------------------------------------------------------------------------
\section{Infrared color--color diagnostics from IRAC photometry}
% ---------------------------------------------------------------------------
\label{sec:irac_diagnostics}

To characterize the SEDs of the final sample in the MIR, we computed the IRAC spectral index, $\alpha_{\rm IRAC}$, defined as

\begin{equation}
\alpha_{\rm IRAC}
=
\frac{d\log(\lambda F_{\lambda})}{d\log(\lambda)}.
\end{equation}
We used the photometry in the four IRAC bands centered at 3.6, 4.5, 5.8, and 8.0~$\mu$m. The Vega magnitudes were first converted into flux densities per unit frequency according to $F_{\nu}=F_{\nu,0}\,10^{-0.4m}$, where the adopted IRAC zero-point flux densities are $F_{\nu,0}=(280.9,\,179.7,\,115.0,\,64.13)$ Jy for the 3.6, 4.5, 5.8, and 8.0~$\mu$m bands, respectively, following \citet{Reach2005}. The resulting $F_{\nu}$ values were converted to flux densities per unit wavelength using $F_{\lambda}=(c/\lambda^{2})F_{\nu}$.

For each source, we then performed a linear least-squares fit of the form $\log(\lambda F_{\lambda})=\alpha_{\rm IRAC}\log(\lambda)+\beta$ using the four IRAC measurements. Only sources with valid photometry in all four bands were included. In the present analysis, $\alpha_{\rm IRAC}$ is interpreted primarily as a descriptor of the infrared SED morphology rather than as a unique determination of the physical evolutionary stage.

% -------------------------------------------------------------------------- FIGURE 6
\begin{figure*}[ht!]
    \centering
    \includegraphics[width=0.85\columnwidth]{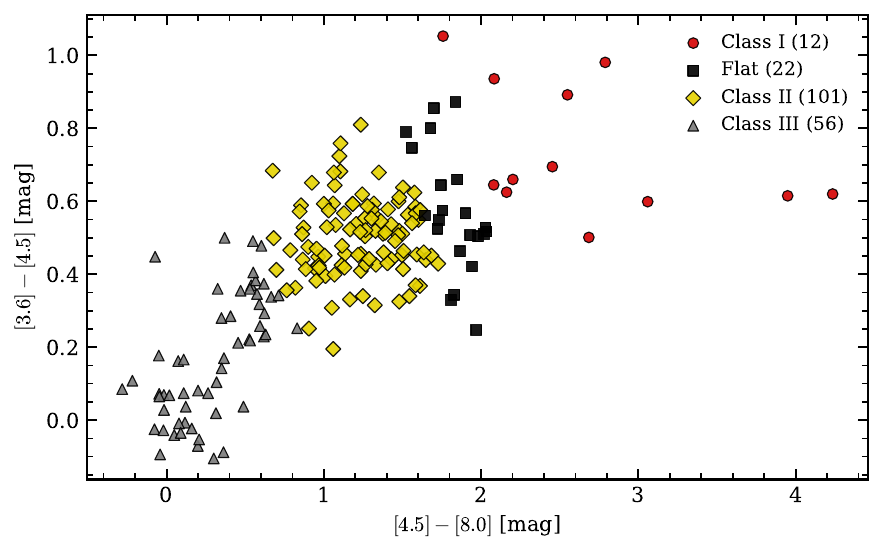}
    \caption{IRAC color--color diagram for the 191 MIR-excess sources, showing $[4.5]-[8.0]$ along the horizontal axis and $[3.6]-[4.5]$ along the vertical axis. Symbols denote the IRAC spectral-index classes derived from the computed $\alpha_{\mathrm{IRAC}}$ values following the classification scheme of \citet{Greene1994}: Class I sources (red circles), Flat-spectrum sources (black squares), Class II sources (yellow diamonds), and Class III sources (gray triangles). The corresponding numbers of objects are 12, 22, 101, and 56, respectively.}    
    \label{fig:irac_ccd}
\end{figure*}
% -------------------------------------------------------------------------- end FIGURE 6

The sources were assigned to the IRAC SED classes defined by \citet{Greene1994}: Class I for $\alpha_{\rm IRAC}>0.3$, Flat-spectrum for $-0.3\leq\alpha_{\rm IRAC}\leq0.3$, Class II for $-1.6<\alpha_{\rm IRAC}<-0.3$, and Class III for $\alpha_{\rm IRAC}\leq-1.6$. Application of these criteria to the 191 MIR-excess sources yields 12 Class I objects, 22 Flat-spectrum sources, 101 Class II objects, and 56 Class III objects. These populations correspond to 6.3\%, 11.5\%, 52.9\%, and 29.3\% of the sample, respectively.

Figure~\ref{fig:irac_ccd} presents the IRAC color--color diagram, with $[4.5]-[8.0]$ along the horizontal axis and $[3.6]-[4.5]$ along the vertical axis. The four spectral-index classes occupy preferential, although partially overlapping, regions of the diagram. Class III sources are concentrated toward the lowest IRAC colors, whereas Class II objects define the dominant locus of the sample at intermediate colors. Flat-spectrum sources are located mainly between the Class II and Class I populations, while the Class I objects preferentially exhibit the largest $[4.5]-[8.0]$ colors. The overall distribution therefore traces a continuous progression toward increasing MIR excess.

In the context of the IRAC spectral-index classification defined above, the representative SEDs shown in Fig.~\ref{fig:sed_mosaic_1} illustrate the diversity of infrared properties observed in the final sample and cover the range of MIR characteristics inferred from the VOSA analysis, from sources exhibiting strong MIR excess to objects whose SEDs remain largely compatible with predominantly photospheric emission, within the observational uncertainties, up to the IRAC bands. Together, these examples demonstrate the complementary roles of SED morphology and the IRAC spectral index in characterizing the infrared-excess sources.

The representative examples shown in Fig.~\ref{fig:sed_mosaic_1} illustrate characteristic SED morphologies observed within the sample. The upper panels (A and B) display the strongest MIR excesses and correspond to the highest values of $\alpha_{\rm IRAC}$. The middle panels (C and D) represent intermediate levels of infrared excess, whereas the lower panels (E and F) exhibit weaker excesses, with observed fluxes remaining closer to the fitted photospheric models over most of the IRAC wavelength range despite satisfying the adopted MIR color-selection criteria.

Although the representative SEDs provide a useful qualitative view of the detected infrared excesses, the physical interpretation of individual objects should not be based solely on SED morphology. Similar observational signatures can arise from different physical configurations, including variations in circumstellar material, disk geometry, evolutionary stage, or contamination by unrelated astrophysical sources.

We therefore adopt a cautious interpretation of the physical nature of the MIR-excess sources, particularly for the less embedded objects, since IRAC spectral index, color-based classifications, and SED morphology may, in some cases, encompass sources of different physical origins.

The predominance of Class II sources indicates that a substantial fraction of the MIR-excess sample has SED slopes consistent with circumstellar-disk emission rather than with the more strongly rising infrared continua commonly associated with embedded sources. The Class I and Flat-spectrum populations nevertheless demonstrate that the sample also includes objects with stronger MIR excesses. Conversely, the significant Class III population occupies the low-color region of the diagram and displays comparatively weak IRAC excesses. The predominance of Class II sources may also reflect the longer duration of the Class II evolutionary stage compared with the shorter-lived Class I and Flat-spectrum stages.

The partial overlap between adjacent classes is expected because the \citet{Greene1994} spectral-index scheme and classifications based on empirical
IRAC color cuts do not represent identical diagnostics. In particular, the color criteria developed by \citet{Gutermuth2009} use combinations of IRAC
colors and contaminant-rejection conditions, whereas the classes adopted here are determined exclusively from the slope of $\lambda F_{\lambda}$ across the
four IRAC bands. The classifications shown in Fig.~\ref{fig:irac_ccd} should therefore be interpreted as spectral-index classes derived from the IRAC SED slope rather than as independent color-selected evolutionary assignments.

Neither IRAC colors nor spectral indices alone can uniquely establish the physical origin of the infrared excess. Consistent with the variability analysis and
the cross-matches with external catalogs, part of the sample may include evolved stars or binary systems in which the infrared excess arises from circumstellar envelopes or companion-related effects rather than from
protoplanetary disks.

% SECTION 7 ---------------------------------------------------------------------------
\section{Variability properties from Gaia and NEOWISE}
% ---------------------------------------------------------------------------
\label{sec:variability}

We searched for time-resolved MIR photometry from \textit{NEOWISE} \citep{Mainzer2014} for all 191 MIR-excess sources. The decade-scale baseline and repeated coverage in W1 (3.4\,$\mu$m) and W2 (4.6\,$\mu$m) enable a direct assessment of variability linked to circumstellar environments, including changes in accretion, inner-disk thermal emission, and variable extinction along the line of sight.

% SUBSECTION 7.1 ---------------------------------------------------------------------------
\subsection{Gaia variability flags}
% ---------------------------------------------------------------------------
\label{sec:gaia_variability}

Photometric variability information from \textit{Gaia} was used as an auxiliary diagnostic to assess whether the MIR excesses identified in our sample may be influenced by time-dependent phenomena. We used the \texttt{phot\_variable\_flag} provided in the \textit{Gaia} DR3 main source table, which indicates the presence of statistically significant photometric variability detected by Gaia’s internal pipelines \citep{GaiaCollaboration2023}.

This flag is derived from dedicated variability processing chains within the Gaia mission and does not encode information on the physical origin or variability class of the source. In this work, we therefore do not attempt to classify variability types based on Gaia data alone. Instead, the \texttt{phot\_variable\_flag} is used solely as a consistency check to identify sources for which photometric variability may contribute to, or modulate, the observed MIR excess.

Out of the 191 MIR-excess sources analyzed, 29 objects (approximately 15\%) are flagged as variable in Gaia DR3, while the remaining 162 sources (about 85\%) have no variability information available. This distribution reflects the reduced sensitivity of Gaia to detect variability in faint, highly reddened, or crowded regions such as those probed by the VVVX survey.

To investigate the nature of the variable subset, we cross-matched the 29 Gaia-variable sources with external variability and classification catalogs. Fourteen sources are classified as YSOs in the SPICY catalog \citep{Kuhn2021}. Although a total of 20 positional matches are found between the variable sample and the SPICY catalog, only 14 are classified as bona fide YSOs (Classes~I--III), while the remaining matches correspond to field stars in the SPICY classification scheme and are therefore not considered YSOs in this work.

Using Gaia DR3 variability catalogs, we further identify 12 sources as long-period variables (LPVs; \citealt{Lebzelter2023}) and 2 sources as eclipsing binaries (EBs; \citealt{Mowlavi2023}). Four of these LPV candidates also appear in the SPICY catalog, consistent with the fact that some YSOs can exhibit long-term photometric variability.

These results highlight the heterogeneous nature of the variable MIR-excess sources. While a significant fraction is consistent with a YSO interpretation, a comparable fraction corresponds to evolved stars or binary systems, whose infrared excess may arise from circumstellar envelopes or companion-related effects rather than from protoplanetary disks. For this reason, variability information is used here in a contextual sense only, and no sources are excluded from the analysis based on Gaia variability diagnostics.

All external variability classifications reported in this work are taken directly from published surveys, namely the Gaia DR3 variability catalogs and the SPICY catalog, through positional cross-matching. No independent variability classification or period determination was performed as part of this study. These classifications are provided solely for context and should not be interpreted as results derived from or independently validated by the present analysis.

For comparison with previous all-sky infrared variability surveys, we cross-matched our final catalog of 191 MIR-excess sources with the WISE variability catalog of \citet{Petrosky2021} using a matching radius of $1^{\prime\prime}$. The cross-match yielded 35 counterparts, corresponding to 18.3\% of the sample, indicating that most of the MIR-excess sources identified in this work are not represented in that catalog. Among the matched sources, three are externally classified as YSOs in the SPICY catalog, three as LPVs in the Gaia DR3 variability catalog, and one as an EB in Gaia DR3, while the remaining 28 sources have no external variability classification. This limited overlap further supports the conclusion that MIR-excess selection and variability-based surveys identify largely complementary source populations.

% SUBSECTION 7.2 ---------------------------------------------------------------------------
\subsection{MIR light curves from NEOWISE}
% ---------------------------------------------------------------------------
\label{sec:neowise_lc}

NEOWISE single-exposure photometry \citep{Mainzer2014} suitable for light-curve analysis is available for 172 of the 191 MIR-excess sources in our final sample. For an additional 12 sources, only one or two NEOWISE L1b measurements were found within the adopted search radius, preventing the construction of meaningful light curves, while no valid NEOWISE L1b detections are available for the remaining seven sources.

% -------------------------------------------------------------------------- FIGURE 7
\begin{figure*}[ht!]
\centering
\includegraphics[width=0.95\columnwidth]{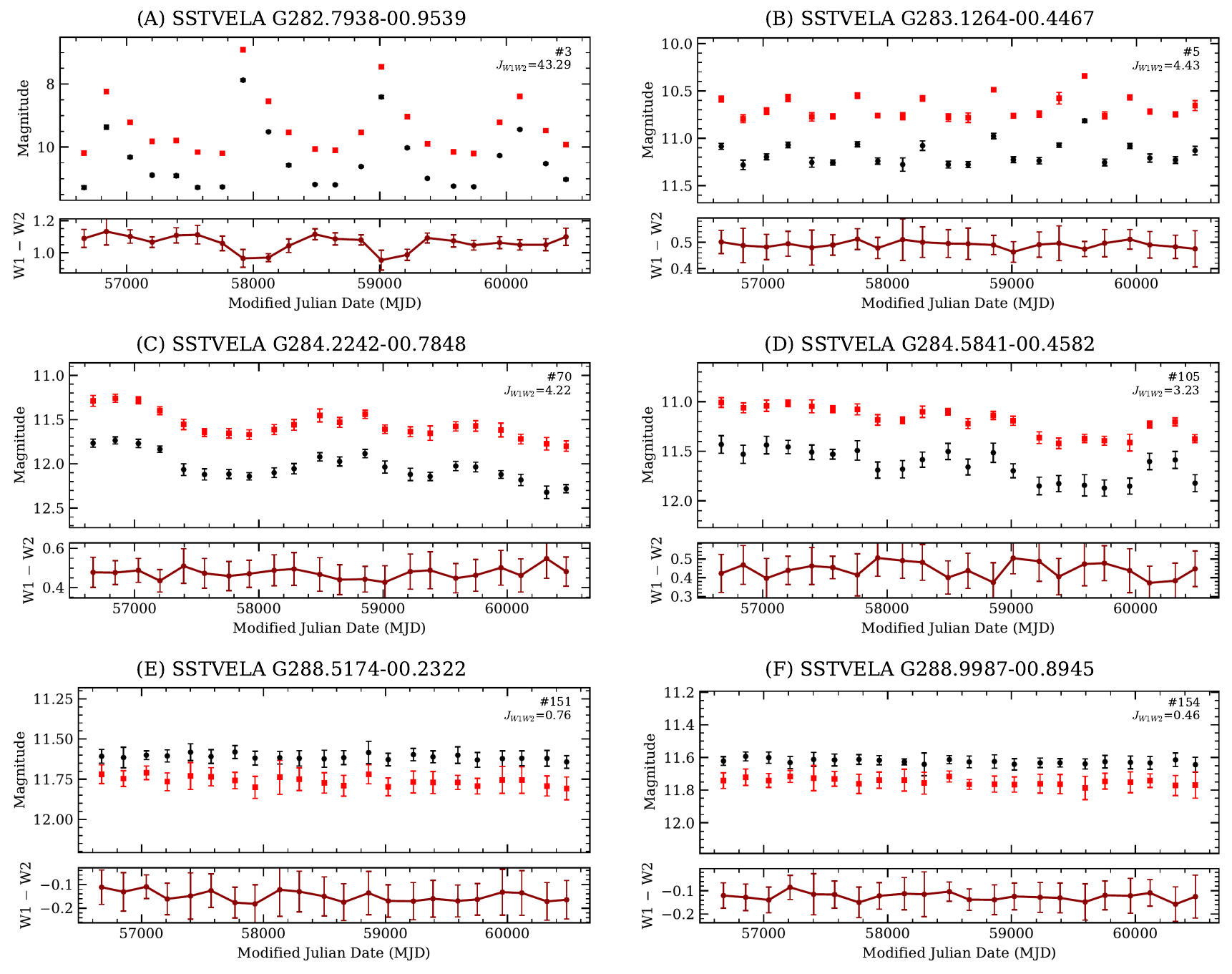}
\caption{Representative NEOWISE light curves of six MIR-excess sources from the final sample, arranged in decreasing order of the Stetson variability index $J_{\mathrm{W1W2}}$: (A) \#3, (B) \#5, (C) \#70, (D) \#105, (E) \#151, and (F) \#154. In each case, the upper panel shows the epoch-averaged W1 (3.4 $\mu$m; black circles) and W2 (4.6 $\mu$m; red squares) magnitudes as a function of modified Julian date, while the lower panel shows the corresponding $W1-W2$ color variation. The plotted values are the median magnitudes measured within each NEOWISE visit, and the error bars represent the standard deviation of the single-exposure measurements in that epoch. The annotations indicate the source identifier and the Stetson variability index $J_{\mathrm{W1W2}}$. Only sources with sufficient temporal sampling are included.}
\label{fig:neowise_lc}
\end{figure*}
% -------------------------------------------------------------------------- end  FIGURE 7

The individual measurements were retrieved from the \textit{ALLWISE} Multi-Epoch Photometry Table via the NASA/\allowbreak IPAC Infrared Science Archive (IRSA). For each source, we adopted the profile-fit magnitudes (\texttt{w1mpro} and \texttt{w2mpro}), which are the recommended photometric measurements for unresolved sources in
the WISE/NEOWISE database. The light-curve construction and subsequent time-series analysis followed the methodology described by \citet{Morris2025}.

To ensure reliable photometry, conservative quality criteria were applied. We required a minimum signal-to-noise ratio of five in both bands ($\texttt{w1snr} \geq 5$ and $\texttt{w2snr} \geq 5$) and adopted the goodness-of-fit thresholds proposed by \citet{Koenig2014} to reject measurements with poor point-spread-function fits:

\begin{equation}
\texttt{w1rchi2} < \frac{\texttt{w1snr} - 3}{7},
\end{equation}
\begin{equation}
\texttt{w2rchi2} < 0.1 \times \texttt{w2snr} - 0.3.
\end{equation}
Measurements failing any of these conditions were discarded.

The remaining single-exposure photometry was grouped into discrete observational epochs corresponding to individual NEOWISE visits. A new epoch was defined whenever the temporal separation between consecutive measurements exceeded 30 days, reproducing the typical cadence of the mission. For each epoch, we computed the median W1 and W2 magnitudes, while the associated uncertainty was estimated as the standard deviation of the measurements within that epoch; catalog-provided photometric errors were not used at this stage.

Figure~\ref{fig:neowise_lc} presents representative NEOWISE light curves illustrating the diversity of MIR variability observed in the final sample. The panels are arranged in decreasing order of the Stetson variability index $J_{\mathrm{W1W2}}$, from the most strongly correlated variable source to the least variable example. The examples include objects exhibiting nearly constant brightness, long-term monotonic variations, and larger changes in both brightness and $W1-W2$ color. These representative light curves emphasize that the amplitude and temporal behavior of the MIR variability are heterogeneous across the sample.

To quantify correlated MIR variability, we computed the Stetson $J$ index \citep{Stetson1996} using contemporaneous $W1$ and $W2$ measurements. Among the 191 sources, 171 provide valid paired measurements suitable for this calculation. Within this subsample, 83.0\% exhibit $J > 0.5$, 64.3\% exceed $J > 1.0$, and 19.9\% show $J > 3.0$, indicating progressively stronger levels of correlated variability. These thresholds are used here as empirical indicators rather than formal classification criteria.

MIR variability is widespread among the infrared-excess sources analyzed here. All sources with valid NEOWISE time series have $n_{\rm pairs} \ge 3$, except for a single object with $n = 2$, for which the Stetson index is not considered reliable.

Cross-identification with the Gaia DR3 variability catalogs reveals that the variable subsample includes 12 LPVs and 2 EBs. Additional parameters available for the two EBs indicate orbital periods of 14.052573 and 10.769329 days, photometric amplitudes of $A_G=0.04$ and $0.30$ mag, and estimated component temperatures of approximately 4800/5000 K and 5700/7700 K, respectively. These properties are summarized in Table~\ref{tab:ebs}.

\begin{table}[ht]
\centering
\caption{Gaia DR3 eclipsing binaries identified in the MIR-excess sample. The first column lists the source number (\#) adopted throughout this work and in the catalog presented in Appendix~A.}
\label{tab:ebs}
\begin{tabular}{cccccc}
\hline
\# & Gaia DR3 ID & 2MASS ID & Period (d) & $A_G$ (mag) & $T_1/T_2$ (K) \\
\hline
131 & 5255398816094663680 & J10290807$-$5831143 & 14.052573 & 0.04 & 4800/5000 \\
142 & 5338255332322386304 & J10520579$-$6024382 & 10.769329 & 0.30 & 5700/7700 \\
\hline
\end{tabular}
\end{table}

A combined analysis of VVVX and NEOWISE time-series photometry is beyond the scope of the present work and will be presented in a forthcoming study.

% SECTION 8 ---------------------------------------------------------------------------
\section{Conclusions}
% ---------------------------------------------------------------------------
\label{sec:conclusions}
We have constructed a homogeneous catalog of 191 MIR-excess sources in the Vela--Carina region covered by the VVVX survey by combining NIR and MIR photometry with SED modeling and complementary variability information from \textit{Gaia} and \textit{NEOWISE}. Starting from a matched VVVX--GLIMPSE--DECaPS catalog of 249\,719 sources, we identified 1\,512 MIR-excess candidates, of which 191 satisfy $\mathrm{S/N}>10$ in all 12 photometric bands and constitute the final sample analyzed in this work. The sample was selected using simple MIR color criteria and subsequently characterized through independent SED analysis.

The \texttt{VOSA} SED fitting confirms that the catalog spans a broad range of infrared properties, from sources with strong MIR excesses to objects exhibiting only moderate or marginal infrared excesses. The predominance of Class II sources indicates that circumstellar-disk emission is the most common infrared characteristic of the sample, while the Class I and Flat-spectrum populations demonstrate the presence of objects with stronger MIR excesses. The predominance of Class II sources may also reflect the longer duration of this evolutionary stage compared with the shorter-lived Class I and Flat-spectrum stages. For the 181 sources with available photogeometric distances, the adopted distances are found to be statistically consistent with the corresponding geometric distance estimates for almost the entire sample, supporting their use in the physical characterization of the selected sources.

A representative subset of SEDs illustrates this diversity of infrared properties, ranging from strong MIR excesses to comparatively weak excesses. Together with the IRAC spectral index, the SED morphology provides complementary diagnostics for characterizing the infrared-excess population and emphasizes the heterogeneous nature of the catalog.

Complementary variability information shows that a minority of the sources in our sample are flagged as photometrically variable by \textit{Gaia}, while the time-resolved \textit{NEOWISE} photometry available for the analyzed subsample indicates the presence of MIR variability, ranging from modest to more pronounced changes depending on the source. Cross-identification with external variability and classification catalogs indicates that the variable subsample includes a mixture of YSOs, evolved LPVs, and EBs. This suggests that, while a significant fraction of the MIR-excess sources is consistent with a YSO interpretation, other objects may host infrared excess arising from circumstellar envelopes or companion-related effects rather than from protoplanetary disks.

The limited overlap with variability-selected YSO catalogs highlights the complementary nature of MIR-excess and variability-based selection techniques. This catalog therefore provides a robust and largely independent sample of infrared-excess sources in the Vela--Carina region of the Galactic plane, with homogeneous multiband photometry, SED characterization, distance information, and variability diagnostics, offering a valuable benchmark for future spectroscopic and time-domain studies of circumstellar environments.

\section*{Acknowledgements}

Data products from the VVVX Survey were obtained with the VISTA telescope at the ESO Paranal Observatory (programme ID~198.B-2004). 
D.M. acknowledges support from the Center for Astrophysics and Associated Technologies (CATA) through the ANID BASAL grants ACE210002 and FB210003, and Fondecyt Project No.~1220724. 

The analysis benefited from VOSA, developed under the Spanish Virtual Observatory project\footnote{\url{https://svo.cab.inta-csic.es}}, funded by 
MCIN\slash AEI\slash 10.13039\slash 501100011033 through grant PID2020-112949GB-I00, with partial updates supported by the EU Horizon~2020 Programme under Grant Agreement No.~776403 (EXOPLANETS-A). Extensive use was made of TOPCAT\footnote{\url{http://www.star.bris.ac.uk/~mbt/topcat/}} \citep{Taylor2005} for the exploration and visual inspection of astronomical catalogs.

Photometric data from the DECaPS survey \citep{Saydjari2023}, obtained with DECam at Cerro Tololo Inter-American Observatory, were incorporated. Additional MIR measurements were retrieved from the \textit{Spitzer}/IRAC GLIMPSE Vela--Carina Archive provided by the NASA/IPAC Infrared Science Archive (IRSA)\footnote{\url{https://irsa.ipac.caltech.edu}}; the \textit{Spitzer Space Telescope} is operated by the Jet Propulsion Laboratory, California Institute of Technology, under contract with NASA.

Astrometric and photometric parameters were taken from the ESA \textit{Gaia} mission, processed by the \textit{Gaia} Data Processing and Analysis Consortium (DPAC), with funding provided by the institutions participating in the Gaia Multilateral Agreement. Distance estimates were obtained from the BJ21 catalog, accessed via the VizieR service at the CDS, Strasbourg.

This study also made use of CDS VizieR catalog access tools and Virtual Observatory services compliant with IVOA standards, including TAP queries executed through TOPCAT. We acknowledge the availability of the three-dimensional dust map of Z25 used in this analysis.

The author acknowledges institutional support from Instituto Federal Catarinense (IFC), which granted research leave to conduct postdoctoral work at Universidade Federal de Santa Catarina (UFSC).

% --- Appendix ---
\clearpage
\onecolumn
\appendix
\setcounter{table}{0}

\section{Catalog table}
\label{app:catalog}

\FloatBarrier
\begin{sidewaystable*}
\caption{Photometric, astrometric, and physical properties of the MIR-excess candidates.}
\label{tab:pms_properties}
\centering
\tiny
\setlength{\tabcolsep}{0.8pt}
\renewcommand{\arraystretch}{0.92}
\begin{tabular}{c l c c c c c c c c c c c c c c c c c c c c c c c c c c}
\hline\hline
\# & GLIMPSE ID & RA & Dec & $g$ & $r$ & $i$ & $z$ & $Y$ & $J$ & $H$ & $K_s$ & $[3.6]$ & $[4.5]$ & $[5.8]$ & $[8.0]$ & $\alpha_{\rm IRAC}$ & Class & Model & $T_{\rm eff}$ & $\log g$ & $L_{\rm bol}$ & $V_{\rm gfb}$ & $A_V$ & $D_{\rm photogeo}$ & RUWE & $n_{\rm pairs}$ & $J_{\rm W1W2}$ \\
 &  & (deg) & (deg) & (mag) & (mag) & (mag) & (mag) & (mag) & (mag) & (mag) & (mag) & (mag) & (mag) & (mag) & (mag) &  &  &  & (K) & (dex) & ($L_\odot$) &  & (mag) & (pc) &  &  &  \\
\hline
1 & G282.8067-00.5030 & 153.55642 & -57.09996 & $17.299(4)$ & $16.281(4)$ & $15.779(5)$ & $15.387(5)$ & $15.195(5)$ & $13.930(7)$ & $13.096(4)$ & $12.274(6)$ & $11.016(43)$ & $10.562(61)$ & $10.019(49)$ & $8.971(38)$ & -0.49 & Class II & BSC & $6700(50)$ & $3.00(25)$ & $15.372(12)$ & 18.5 & 3.24 & $2320_{-215}^{+284}$ & 1.00 & 520 & 5.18 \\
2 & G282.7102-00.3679 & 153.54987 & -56.93395 & $22.332(22)$ & $19.392(7)$ & $17.701(6)$ & $16.431(5)$ & $15.895(5)$ & $13.796(5)$ & $12.470(10)$ & $11.788(6)$ & $11.192(50)$ & $10.912(57)$ & $10.709(62)$ & $10.563(45)$ & -2.13 & Class III & BSC & $3800(50)$ & $3.00(25)$ & $40.338(58)$ & 9.1 & 5.65 & $4273_{-691}^{+750}$ & 1.07 & 517 & 1.88 \\
3 & G282.7938-00.9539 & 153.06394 & -57.46411 & $23.187(62)$ & $20.375(13)$ & $18.472(9)$ & $17.268(8)$ & $16.804(8)$ & $14.276(5)$ & $12.903(4)$ & $11.810(6)$ & $10.565(37)$ & $9.818(50)$ & $9.134(34)$ & $8.259(36)$ & -0.20 & Flat & BS & $2600(50)$ & $4.00(25)$ & $5.905(21)$ & 44.2 & 3.67 & $2669_{-792}^{+1016}$ & 1.04 & 518 & 43.29 \\
4 & G283.0789-00.2649 & 154.21671 & -57.05505 & $20.564(9)$ & $18.131(4)$ & $16.721(5)$ & $15.777(5)$ & $15.362(5)$ & $13.587(8)$ & $12.738(5)$ & $12.100(7)$ & $11.228(51)$ & $10.873(59)$ & $10.626(69)$ & $10.402(47)$ & -1.91 & Class III & KU03 & $4500(125)$ & $5.00(25)$ & $62.571(57)$ & 5.9 & 4.93 & $5067_{-729}^{+1583}$ & 1.08 & 128 & 2.97 \\
5 & G283.1264-00.4467 & 154.10298 & -57.23255 & $19.510(5)$ & $17.330(4)$ & $16.144(5)$ & $15.232(5)$ & $14.852(5)$ & $13.288(10)$ & $12.465(6)$ & $11.870(7)$ & $11.150(36)$ & $10.790(64)$ & $10.411(65)$ & $10.465(83)$ & -2.03 & Class III & BSC & $5100(50)$ & $3.00(25)$ & $73.881(60)$ & 13.8 & 4.94 & $4161_{-468}^{+470}$ & 1.00 & 501 & 4.43 \\
6 & G282.9066-00.3266 & 153.89058 & -57.01010 & $20.431(7)$ & $18.033(5)$ & $16.704(5)$ & $15.912(5)$ & $15.527(5)$ & $13.882(8)$ & $13.050(5)$ & $12.348(7)$ & $11.198(48)$ & $10.828(48)$ & $10.445(65)$ & $10.290(50)$ & -1.78 & Class III & BSC & $5100(50)$ & $3.00(25)$ & $49.064(39)$ & 6.5 & 5.21 & $4293_{-493}^{+615}$ & 0.90 & 509 & 7.81 \\
7 & G286.6973-00.7810 & 159.56344 & -59.38813 & $22.801(67)$ & $20.550(18)$ & $19.073(13)$ & $18.149(7)$ & $17.677(10)$ & $15.520(10)$ & $13.839(6)$ & $12.320(7)$ & $10.619(45)$ & $9.940(52)$ & $9.293(42)$ & $8.590(63)$ & -0.51 & Class II & BS & $2600(50)$ & $4.00(25)$ & $23.834(187)$ & 63.9 & 3.63 & $7525_{-2145}^{+4477}$ & 1.09 & 505 & 1.74 \\
8 & G287.9328-00.4454 & 162.02362 & -59.67612 & $18.327(6)$ & $16.715(5)$ & $16.520(6)$ & $16.063(5)$ & $15.853(5)$ & $14.223(7)$ & $13.340(10)$ & $12.523(7)$ & $11.811(47)$ & $11.471(70)$ & $10.989(80)$ & $10.223(49)$ & -1.00 & Class II & BSC & $4900(50)$ & $3.50(25)$ & $5.755(9)$ & 18.7 & 2.29 & $2675_{-356}^{+397}$ & 1.06 & 469 & 4.14 \\
9 & G282.3240-00.8120 & 152.50250 & -57.07863 & $19.184(5)$ & $17.008(5)$ & $15.864(6)$ & $15.157(7)$ & $14.855(6)$ & $13.247(6)$ & $12.372(13)$ & $11.572(8)$ & $10.180(39)$ & $9.580(46)$ & $8.983(39)$ & $8.105(54)$ & -0.46 & Class II & BSC & $6300(50)$ & $3.50(25)$ & $24.793(14)$ & 11.0 & 5.56 & $1737_{-144}^{+155}$ & 0.97 & \dots & \dots \\
10 & G283.0305-00.1440 & 154.26605 & -56.92758 & $20.590(17)$ & $17.822(8)$ & $16.269(7)$ & $15.283(7)$ & $14.846(7)$ & $12.970(11)$ & $12.033(6)$ & $11.378(8)$ & $10.721(69)$ & $10.492(90)$ & $10.166(56)$ & $9.873(38)$ & -1.85 & Class III & BSC & $4600(50)$ & $3.00(25)$ & $129.958(145)$ & 4.2 & 5.69 & $4861_{-1097}^{+1147}$ & 0.96 & 29 & 0.23 \\
11 & G283.1452-00.2896 & 154.29297 & -57.11236 & $21.178(14)$ & $18.923(7)$ & $17.696(7)$ & $18.209(11)$ & $16.430(7)$ & $14.856(14)$ & $13.962(12)$ & $13.442(8)$ & $12.379(54)$ & $11.844(88)$ & $11.543(76)$ & $10.769(52)$ & -1.05 & Class II & BSC & $3800(50)$ & $3.00(25)$ & $0.440(2)$ & 55.0 & 1.63 & $1523_{-261}^{+327}$ & 1.32 & \dots & \dots \\
12 & G283.1277-00.8353 & 153.70142 & -57.55527 & $19.971(10)$ & $17.674(6)$ & $16.431(10)$ & $15.520(10)$ & $15.118(10)$ & $13.453(6)$ & $12.640(7)$ & $12.059(8)$ & $11.525(87)$ & $11.025(93)$ & $10.817(80)$ & $10.656(66)$ & -1.89 & Class III & BSC & $4600(50)$ & $3.00(25)$ & $41.659(45)$ & 1.9 & 4.62 & $3937_{-463}^{+453}$ & 0.99 & \dots & \dots \\
13 & G282.7902-00.6530 & 153.37511 & -57.21436 & $23.381(66)$ & $21.019(18)$ & $19.659(12)$ & $18.506(8)$ & $18.006(10)$ & $16.073(7)$ & $14.681(7)$ & $13.453(8)$ & $11.714(53)$ & $11.203(60)$ & $10.776(56)$ & $9.827(43)$ & -0.70 & Class II & BS & $2600(50)$ & $4.00(25)$ & $1.131(10)$ & 49.9 & 2.64 & $2658_{-698}^{+1613}$ & 1.09 & 522 & 2.46 \\
14 & G282.2191-00.9186 & 152.23143 & -57.10498 & $23.584(96)$ & $20.445(13)$ & $18.503(8)$ & $17.163(6)$ & $16.612(7)$ & $14.382(6)$ & $12.697(5)$ & $11.336(8)$ & $9.574(61)$ & $8.890(49)$ & $8.570(39)$ & $8.215(26)$ & -1.34 & Class II & BSC & $3500(50)$ & $3.00(25)$ & $25.236(171)$ & 26.0 & 5.83 & $3818_{-928}^{+2011}$ & 1.08 & 518 & 1.00 \\
15 & G282.5237-00.5305 & 153.09924 & -56.96228 & $20.512(11)$ & $18.706(7)$ & $17.740(8)$ & $16.935(8)$ & $16.594(8)$ & $15.108(6)$ & $14.292(7)$ & $13.655(8)$ & $12.627(56)$ & $12.042(62)$ & $11.483(101)$ & $10.460(35)$ & -0.37 & Class II & BSC & $4700(50)$ & $3.00(25)$ & $8.485(11)$ & 7.5 & 3.80 & $3977_{-922}^{+1624}$ & 1.03 & 511 & 1.09 \\
16 & G282.1662-00.4747 & 152.62170 & -56.71206 & $23.237(48)$ & $19.996(7)$ & $18.061(6)$ & $16.901(5)$ & $16.467(6)$ & $14.420(5)$ & $12.907(6)$ & $11.630(8)$ & $9.589(37)$ & $8.717(44)$ & $7.941(36)$ & $6.880(27)$ & 0.26 & Flat & BSC & $4000(50)$ & $3.00(25)$ & $16.132(50)$ & 16.4 & 6.54 & $2666_{-572}^{+666}$ & 1.03 & 523 & 2.52 \\
17 & G283.9700-00.1362 & 155.72592 & -57.43435 & $22.446(31)$ & $20.438(12)$ & $19.377(13)$ & $18.507(9)$ & $18.110(13)$ & $16.259(13)$ & $14.617(9)$ & $13.331(8)$ & $11.767(49)$ & $11.248(72)$ & $10.742(56)$ & $9.769(50)$ & -0.56 & Class II & BSC & $3900(50)$ & $3.00(25)$ & $5.906(28)$ & 29.3 & 4.42 & $4676_{-941}^{+1279}$ & 1.09 & 225 & 2.98 \\
18 & G286.3903-00.7191 & 159.09956 & -59.18328 & $21.739(23)$ & $19.276(9)$ & $17.899(8)$ & $16.881(8)$ & $16.383(8)$ & $14.259(9)$ & $13.054(8)$ & $12.176(8)$ & $10.554(40)$ & $10.138(48)$ & $9.630(42)$ & $9.195(30)$ & -1.26 & Class II & BSC & $3900(50)$ & $3.00(25)$ & $63.640(149)$ & 16.1 & 5.13 & $6848_{-798}^{+2657}$ & 1.68 & 517 & 3.63 \\
19 & G288.5715-00.2815 & 163.30278 & -59.81551 & $20.662(9)$ & $18.859(7)$ & $18.169(6)$ & $17.451(5)$ & $17.121(6)$ & $15.089(10)$ & $14.117(7)$ & $13.557(8)$ & $12.967(53)$ & $12.623(94)$ & $11.751(87)$ & $10.794(56)$ & -0.25 & Flat & \dots & \dots & \dots & \dots & \dots & \dots & \dots & 1.03 & 328 & 2.18 \\
20 & G288.5746-00.4633 & 163.14871 & -59.98012 & $20.281(8)$ & $18.819(8)$ & $18.285(7)$ & $18.210(7)$ & $17.425(8)$ & $15.428(6)$ & $14.309(7)$ & $13.435(8)$ & $12.258(58)$ & $11.927(87)$ & $11.530(76)$ & $10.761(58)$ & -1.12 & Class II & BSC & $3900(50)$ & $3.00(25)$ & $0.563(6)$ & 48.2 & 0.99 & $2046_{-642}^{+676}$ & 1.43 & 481 & 2.10 \\
21 & G288.6542-00.4515 & 163.30212 & -60.00453 & $18.587(6)$ & $17.694(8)$ & $17.284(5)$ & $16.826(5)$ & $16.602(6)$ & $15.115(8)$ & $14.080(9)$ & $13.143(8)$ & $12.089(59)$ & $11.445(53)$ & $10.876(80)$ & $9.702(38)$ & -0.13 & Flat & BSC & $4900(50)$ & $3.50(25)$ & $7.776(25)$ & 26.1 & 1.96 & $4712_{-915}^{+1339}$ & 1.05 & 479 & 0.85 \\
22 & G288.7985-00.4734 & 163.54290 & -60.08736 & $16.679(6)$ & $15.946(7)$ & $15.779(6)$ & $15.534(6)$ & $15.429(6)$ & $14.456(6)$ & $13.706(7)$ & $12.159(8)$ & $9.461(43)$ & $8.606(48)$ & $7.832(37)$ & $6.908(25)$ & 0.09 & Flat & BSC & $7000(50)$ & $3.00(25)$ & $35.757(196)$ & 28.1 & 2.02 & $4626_{-543}^{+644}$ & 1.03 & 455 & 1.82 \\
23 & G283.0378-00.6993 & 153.70520 & -57.39226 & $23.704(101)$ & $21.187(19)$ & $19.598(12)$ & $18.357(9)$ & $17.839(9)$ & $15.917(8)$ & $14.644(6)$ & $13.574(9)$ & $12.002(53)$ & $11.477(60)$ & $10.749(54)$ & $9.755(40)$ & -0.23 & Flat & BSC & $4300(50)$ & $3.00(25)$ & $8.913(18)$ & 16.3 & 6.56 & $4020_{-858}^{+1000}$ & 1.11 & 503 & 1.96 \\
24 & G282.3923-00.0842 & 153.36394 & -56.52026 & $21.732(17)$ & $20.117(11)$ & $18.395(8)$ & $17.510(7)$ & $17.120(8)$ & $15.519(6)$ & $14.076(5)$ & $13.195(9)$ & $11.914(41)$ & $11.540(70)$ & $11.294(69)$ & $10.921(55)$ & -1.73 & Class III & BSC & $3900(50)$ & $3.00(25)$ & $3.119(6)$ & 16.3 & 4.41 & $2593_{-518}^{+1059}$ & 1.03 & 536 & 2.74 \\
25 & G283.0285-00.6599 & 153.73201 & -57.35452 & $23.818(99)$ & $21.364(20)$ & $19.876(13)$ & $18.746(8)$ & $18.246(10)$ & $16.418(8)$ & $14.990(7)$ & $13.822(9)$ & $12.009(77)$ & $11.505(64)$ & $11.016(72)$ & $9.525(40)$ & -0.03 & Flat & BSC & $4200(50)$ & $3.00(25)$ & $9.280(36)$ & 21.8 & 6.46 & $4795_{-590}^{+908}$ & 1.01 & 493 & 1.49 \\
26 & G283.0500-00.9429 & 153.46857 & -57.60036 & $19.507(7)$ & $17.966(6)$ & $17.303(8)$ & $16.658(7)$ & $16.381(8)$ & $14.669(10)$ & $13.827(7)$ & $13.220(9)$ & $12.198(45)$ & $11.750(64)$ & $11.474(98)$ & $10.786(97)$ & -1.26 & Class II & BSC & $4800(50)$ & $3.50(25)$ & $4.708(6)$ & 12.3 & 3.31 & $2635_{-451}^{+581}$ & 1.07 & 523 & 1.70 \\
27 & G282.6109-00.2472 & 153.52454 & -56.77833 & $23.437(70)$ & $20.891(17)$ & $19.498(12)$ & $18.419(8)$ & $17.984(11)$ & $16.073(11)$ & $14.680(6)$ & $13.493(9)$ & $11.893(50)$ & $11.169(69)$ & $10.700(61)$ & $10.070(44)$ & -0.79 & Class II & BSC & $3800(50)$ & $3.00(25)$ & $1.747(7)$ & 20.0 & 5.02 & $2395_{-650}^{+1575}$ & 0.95 & 506 & 1.49 \\
28 & G282.9680-00.5942 & 153.70774 & -57.26615 & $17.564(5)$ & $16.104(4)$ & $15.284(5)$ & $14.790(4)$ & $14.550(5)$ & $13.178(6)$ & $12.404(6)$ & $11.811(9)$ & $10.758(41)$ & $10.433(54)$ & $9.954(47)$ & $8.954(53)$ & -0.76 & Class II & BS & $3700(50)$ & $3.00(25)$ & $1.987(6)$ & 12.2 & 0.53 & $1422_{-96}^{+120}$ & 1.89 & 505 & 2.62 \\
29 & G282.8471-00.9679 & 153.13051 & -57.50594 & $19.433(11)$ & $17.770(11)$ & $17.323(11)$ & $16.597(7)$ & $16.418(11)$ & $14.583(7)$ & $13.659(6)$ & $13.126(9)$ & $11.978(55)$ & $11.449(89)$ & $11.254(72)$ & $10.430(66)$ & -1.14 & Class II & BSC & $3800(50)$ & $3.00(25)$ & $1.065(4)$ & 18.0 & 1.25 & $1861_{-235}^{+326}$ & 1.14 & 512 & 2.55 \\
30 & G282.8623-00.7984 & 153.33267 & -57.37502 & $22.380(37)$ & $19.725(10)$ & $18.318(7)$ & $17.332(6)$ & $16.917(7)$ & $14.933(6)$ & $13.481(5)$ & $12.524(9)$ & $11.230(53)$ & $10.611(43)$ & $10.003(51)$ & $9.367(24)$ & -0.69 & Class II & BSC & $3500(50)$ & $3.00(25)$ & $5.176(26)$ & 22.9 & 3.24 & $3403_{-730}^{+1294}$ & 1.03 & 511 & 3.19 \\
31 & G282.8680-00.9661 & 153.16468 & -57.51639 & $19.780(7)$ & $17.908(5)$ & $17.074(7)$ & $16.534(7)$ & $16.297(8)$ & $14.733(5)$ & $13.753(6)$ & $13.206(9)$ & $12.374(51)$ & $11.780(101)$ & $11.361(77)$ & $10.462(67)$ & -0.69 & Class II & BS & $3600(50)$ & $3.00(25)$ & $0.675(2)$ & 13.7 & 0.86 & $1630_{-500}^{+973}$ & 2.84 & \dots & \dots \\
32 & G282.5656-00.2183 & 153.48609 & -56.72885 & $23.313(59)$ & $20.598(14)$ & $19.020(9)$ & $17.917(7)$ & $17.461(8)$ & $15.602(6)$ & $14.445(6)$ & $13.528(9)$ & $12.269(68)$ & $11.731(44)$ & $11.414(98)$ & $10.525(30)$ & -0.90 & Class II & BS & $3700(50)$ & $4.00(25)$ & $5.193(13)$ & 13.1 & 5.30 & $3446_{-700}^{+977}$ & 1.21 & 517 & 2.41 \\
33 & G283.6853-00.4412 & 154.97327 & -57.53628 & $23.461(101)$ & $20.703(13)$ & $18.947(8)$ & $17.615(6)$ & $17.027(6)$ & $14.880(11)$ & $13.701(13)$ & $12.951(9)$ & $11.828(48)$ & $11.465(62)$ & $10.996(77)$ & $10.646(95)$ & -1.46 & Class II & BS & $3100(50)$ & $3.00(25)$ & $4.574(16)$ & 20.2 & 3.27 & $3626_{-662}^{+1176}$ & 1.08 & 489 & 7.76 \\
34 & G286.7951-00.5372 & 159.96278 & -59.22255 & $21.652(18)$ & $20.073(12)$ & $18.030(6)$ & $17.308(5)$ & $16.900(6)$ & $16.179(10)$ & $14.590(12)$ & $13.129(9)$ & $12.062(55)$ & $11.507(66)$ & $11.014(82)$ & $10.170(57)$ & -0.69 & Class II & BSC & $3900(50)$ & $4.50(25)$ & $3.697(14)$ & 32.0 & 3.44 & $3532_{-830}^{+912}$ & 1.02 & 246 & 1.29 \\
35 & G286.8347-00.5507 & 160.01791 & -59.25352 & $21.969(20)$ & $20.103(10)$ & $18.882(7)$ & $17.776(5)$ & $17.303(6)$ & $15.233(8)$ & $13.948(8)$ & $13.086(9)$ & $11.667(72)$ & $11.023(63)$ & $10.579(68)$ & $9.954(42)$ & -0.91 & Class II & BSC & $3800(50)$ & $3.00(25)$ & $5.144(22)$ & 23.1 & 4.60 & $3285_{-859}^{+1034}$ & 1.17 & 478 & 2.29 \\
36 & G288.5194-00.5429 & 162.97929 & -60.02709 & $17.929(5)$ & $17.201(5)$ & $16.826(5)$ & $16.493(4)$ & $16.340(5)$ & $14.853(7)$ & $13.885(8)$ & $12.927(9)$ & $11.475(62)$ & $10.900(43)$ & $10.396(63)$ & $9.145(28)$ & -0.20 & Flat & BSC & $6000(50)$ & $3.00(25)$ & $8.361(32)$ & 31.2 & 2.01 & $3890_{-659}^{+1021}$ & 0.99 & 419 & 2.70 \\
37 & G288.7509-00.8753 & 163.10108 & -60.42758 & $20.037(6)$ & $17.945(5)$ & $16.734(4)$ & $15.965(4)$ & $15.616(4)$ & $13.999(6)$ & $13.197(8)$ & $12.564(9)$ & $11.649(57)$ & $11.427(81)$ & $11.053(60)$ & $10.901(49)$ & -1.95 & Class III & BSC & $3800(50)$ & $4.50(25)$ & $33.414(91)$ & 10.4 & 2.73 & $6390_{-1463}^{+1136}$ & 1.00 & \dots & \dots \\
38 & G288.7029-00.1593 & 163.64387 & -59.76291 & $21.022(14)$ & $18.368(7)$ & $16.954(6)$ & $15.906(5)$ & $15.424(6)$ & $13.418(7)$ & $12.437(7)$ & $11.702(9)$ & $10.840(49)$ & $10.502(53)$ & $10.112(53)$ & $9.837(40)$ & -1.67 & Class III & BS & $3300(50)$ & $3.00(25)$ & $24.664(68)$ & 14.6 & 2.57 & $4698_{-847}^{+2459}$ & 1.06 & 474 & 7.38 \\
39 & G282.2958-00.7867 & 152.48693 & -57.04172 & $23.591(94)$ & $20.349(10)$ & $18.455(7)$ & $17.240(6)$ & $16.757(6)$ & $14.657(7)$ & $12.973(6)$ & $11.896(10)$ & $10.209(46)$ & $9.615(47)$ & $9.225(46)$ & $8.556(47)$ & -0.98 & Class II & BS & $2600(50)$ & $4.00(25)$ & $4.446(47)$ & 42.8 & 1.63 & $3090_{-702}^{+937}$ & 0.96 & 535 & 4.81 \\
40 & G283.0443-00.8966 & 153.50866 & -57.55899 & $21.504(15)$ & $19.395(7)$ & $18.251(7)$ & $17.272(6)$ & $16.853(7)$ & $15.169(8)$ & $14.112(7)$ & $13.229(10)$ & $11.824(47)$ & $11.296(51)$ & $10.865(59)$ & $10.425(41)$ & -1.25 & Class II & BSC & $3900(50)$ & $3.00(25)$ & $3.855(13)$ & 14.3 & 3.29 & $3288_{-663}^{+958}$ & 1.06 & 504 & 0.54 \\
41 & G283.3457-00.6919 & 154.18773 & -57.55831 & $23.177(56)$ & $20.275(11)$ & $18.692(8)$ & $17.478(7)$ & $16.968(7)$ & $14.996(7)$ & $13.869(5)$ & $13.149(10)$ & $12.227(40)$ & $11.787(63)$ & $11.369(86)$ & $10.921(73)$ & -1.34 & Class II & BS & $3000(50)$ & $4.00(25)$ & $4.717(8)$ & 11.1 & 3.93 & $3348_{-668}^{+1190}$ & 1.08 & 490 & 0.68 \\
42 & G282.8580-00.4748 & 153.66369 & -57.10551 & $20.048(6)$ & $18.215(5)$ & $17.287(5)$ & $16.660(5)$ & $16.376(5)$ & $14.820(6)$ & $13.950(7)$ & $13.211(10)$ & $11.946(50)$ & $11.532(63)$ & $11.220(75)$ & $10.647(56)$ & -1.38 & Class II & BSC & $4500(50)$ & $3.00(25)$ & $2.369(5)$ & 10.7 & 3.21 & $2037_{-316}^{+374}$ & 0.89 & 518 & 4.26 \\
43 & G283.2672-00.1580 & 154.61401 & -57.07018 & $22.454(30)$ & $19.225(7)$ & $17.339(6)$ & $16.070(6)$ & $15.548(6)$ & $13.517(11)$ & $12.350(7)$ & $11.408(10)$ & $10.928(39)$ & $10.586(58)$ & $10.229(44)$ & $9.874(38)$ & -1.63 & Class III & BS & $3400(50)$ & $3.00(25)$ & $12.797(21)$ & 12.4 & 3.85 & $2899_{-391}^{+365}$ & 1.02 & 490 & 0.46 \\
44 & G285.7276-00.3388 & 158.35688 & -58.52288 & $18.985(5)$ & $17.784(5)$ & $17.029(5)$ & $16.486(5)$ & $16.267(6)$ & $14.866(11)$ & $13.954(10)$ & $13.267(10)$ & $12.209(59)$ & $11.734(84)$ & $11.507(89)$ & $10.833(50)$ & -1.32 & Class II & BSC & $4700(50)$ & $3.00(25)$ & $5.923(14)$ & 14.5 & 2.31 & $3626_{-969}^{+1327}$ & 3.63 & 494 & 2.00 \\
45 & G285.9905-00.8962 & 158.25160 & -59.13671 & $18.436(5)$ & $17.331(6)$ & $16.831(5)$ & $16.375(5)$ & $16.156(6)$ & $14.923(9)$ & $14.197(7)$ & $13.341(10)$ & $11.972(45)$ & $11.462(55)$ & $10.860(75)$ & $9.443(33)$ & 0.05 & Flat & BSC & $6300(50)$ & $3.00(25)$ & $17.095(20)$ & 18.0 & 2.89 & $4542_{-1175}^{+1773}$ & 1.04 & 509 & 0.45 \\
46 & G286.6632-00.0692 & 160.17631 & -58.74889 & $18.500(5)$ & $17.389(13)$ & $17.264(5)$ & $16.754(5)$ & $16.528(7)$ & $14.909(7)$ & $14.140(9)$ & $13.244(10)$ & $11.776(46)$ & $11.581(73)$ & $11.015(72)$ & $10.521(66)$ & -1.33 & Class II & BSC & $5200(50)$ & $3.50(25)$ & $3.731(16)$ & 36.8 & 1.87 & $3111_{-447}^{+1017}$ & 1.17 & 425 & 4.28 \\
47 & G286.8039-00.6116 & 159.90752 & -59.29194 & $22.854(49)$ & $20.301(13)$ & $18.687(7)$ & $17.675(6)$ & $17.210(7)$ & $15.231(7)$ & $13.899(5)$ & $13.347(10)$ & $11.756(54)$ & $11.278(71)$ & $10.974(60)$ & $10.675(66)$ & -1.63 & Class III & BSC & $3700(50)$ & $3.00(25)$ & $6.352(20)$ & 13.2 & 4.45 & $3763_{-1132}^{+1852}$ & 1.02 & 383 & 4.93 \\
48 & G286.5638-00.4081 & 159.69090 & -58.99747 & $19.877(14)$ & $18.054(6)$ & $17.103(6)$ & $16.467(7)$ & $16.212(6)$ & $14.676(8)$ & $13.766(10)$ & $13.322(10)$ & $12.593(51)$ & $12.141(83)$ & $11.413(99)$ & $10.644(66)$ & -0.55 & Class II & BS & $3800(50)$ & $3.00(25)$ & $1.959(4)$ & 5.7 & 1.76 & $2385_{-358}^{+502}$ & 0.98 & 478 & 2.42 \\
49 & G282.4449-00.6292 & 152.87718 & -56.99846 & $20.346(13)$ & $19.212(11)$ & $17.140(8)$ & $16.794(8)$ & $16.632(8)$ & $16.068(7)$ & $14.805(8)$ & $13.780(11)$ & $12.583(57)$ & $12.006(83)$ & $11.497(77)$ & $10.700(62)$ & -0.69 & Class II & BSC & $5500(50)$ & $4.00(25)$ & $4.334(7)$ & 33.0 & 3.75 & $3035_{-512}^{+841}$ & 1.02 & 2 & \dots \\
50 & G283.0170-00.6716 & 153.70225 & -57.35769 & $23.173(49)$ & $20.512(11)$ & $18.991(8)$ & $17.795(6)$ & $17.274(6)$ & $15.263(5)$ & $13.782(5)$ & $12.521(11)$ & $10.751(32)$ & $10.240(54)$ & $9.657(33)$ & $8.741(27)$ & -0.53 & Class II & BSC & $3800(50)$ & $3.00(25)$ & $5.875(18)$ & 21.4 & 5.37 & $2926_{-905}^{+933}$ & 1.05 & 499 & 1.44 \\
51 & G283.2806-00.4333 & 154.35394 & -57.30703 & $21.583(16)$ & $19.704(8)$ & $18.695(8)$ & $17.819(7)$ & $17.418(8)$ & $15.726(10)$ & $14.649(8)$ & $13.653(11)$ & $12.500(56)$ & $11.999(68)$ & $10.925(58)$ & $9.314(34)$ & 0.91 & Class I & BSC & $3800(50)$ & $3.00(25)$ & $1.212(5)$ & 18.1 & 2.43 & $2715_{-707}^{+866}$ & 1.04 & 485 & 2.16 \\
52 & G283.4380-00.6164 & 154.40885 & -57.54668 & $23.087(57)$ & $20.756(14)$ & $19.421(12)$ & $18.302(8)$ & $17.835(10)$ & $15.810(7)$ & $14.088(12)$ & $12.742(11)$ & $11.575(40)$ & $11.008(58)$ & $10.409(47)$ & $9.108(29)$ & -0.02 & Flat & BS & $3100(50)$ & $3.00(25)$ & $0.659(5)$ & 45.0 & 1.72 & $2001_{-843}^{+2070}$ & 1.13 & 481 & 4.10 \\
53 & G284.5023-00.9310 & 155.76763 & -58.39053 & $22.846(52)$ & $19.866(11)$ & $18.290(10)$ & $16.991(8)$ & $16.495(8)$ & $14.363(7)$ & $13.296(9)$ & $12.640(11)$ & $11.889(62)$ & $11.441(75)$ & $11.366(93)$ & $11.514(83)$ & -2.47 & Class III & BS & $3200(50)$ & $4.00(25)$ & $19.372(54)$ & 10.2 & 4.48 & $4878_{-638}^{+688}$ & 1.04 & \dots & \dots \\
54 & G283.9653-00.1409 & 155.71379 & -57.43576 & $20.344(9)$ & $18.054(7)$ & $16.943(7)$ & $16.290(7)$ & $15.980(8)$ & $14.705(9)$ & $13.595(10)$ & $12.713(11)$ & $11.557(41)$ & $11.058(59)$ & $10.689(59)$ & $10.377(51)$ & -1.50 & Class II & BSC & $4400(50)$ & $4.50(25)$ & $9.379(16)$ & 11.6 & 3.44 & $3415_{-773}^{+1317}$ & 1.02 & 478 & 2.38 \\
55 & G284.0827-00.9354 & 155.09206 & -58.16741 & $22.477(28)$ & $20.031(9)$ & $18.758(8)$ & $17.809(7)$ & $17.386(8)$ & $15.499(10)$ & $14.240(9)$ & $13.252(11)$ & $11.883(49)$ & $11.459(65)$ & $11.081(81)$ & $10.492(67)$ & -1.26 & Class II & BSC & $3900(50)$ & $3.00(25)$ & $11.786(35)$ & 14.8 & 4.36 & $5540_{-1559}^{+3272}$ & 1.03 & 494 & 2.33 \\
\hline
\end{tabular}
\end{sidewaystable*}

\addtocounter{table}{-1}
\begin{sidewaystable*}
\caption{Photometric, astrometric, and physical properties of the MIR-excess candidates (continued).}
\centering
\tiny
\setlength{\tabcolsep}{0.8pt}
\renewcommand{\arraystretch}{0.92}
\begin{tabular}{c l c c c c c c c c c c c c c c c c c c c c c c c c c c}
\hline\hline
\# & GLIMPSE ID & RA & Dec & $g$ & $r$ & $i$ & $z$ & $Y$ & $J$ & $H$ & $K_s$ & $[3.6]$ & $[4.5]$ & $[5.8]$ & $[8.0]$ & $\alpha_{\rm IRAC}$ & Class & Model & $T_{\rm eff}$ & $\log g$ & $L_{\rm bol}$ & $V_{\rm gfb}$ & $A_V$ & $D_{\rm photogeo}$ & RUWE & $n_{\rm pairs}$ & $J_{\rm W1W2}$ \\
 &  & (deg) & (deg) & (mag) & (mag) & (mag) & (mag) & (mag) & (mag) & (mag) & (mag) & (mag) & (mag) & (mag) & (mag) &  &  &  & (K) & (dex) & ($L_\odot$) &  & (mag) & (pc) &  &  &  \\
\hline
56 & G285.2244-00.7361 & 157.13806 & -58.60658 & $19.591(7)$ & $18.084(6)$ & $17.361(7)$ & $16.857(6)$ & $16.618(7)$ & $15.260(6)$ & $14.466(9)$ & $13.824(11)$ & $13.128(56)$ & $12.702(64)$ & $12.426(86)$ & $11.371(49)$ & -0.88 & Class II & BSC & $5300(50)$ & $4.00(25)$ & $7.806(7)$ & 7.4 & 3.10 & $4047_{-917}^{+1213}$ & 1.05 & 494 & 1.40 \\
57 & G285.5794-00.9229 & 157.53783 & -58.94999 & $22.581(41)$ & $20.001(12)$ & $18.628(8)$ & $17.616(7)$ & $17.169(9)$ & $15.188(7)$ & $13.838(8)$ & $12.863(11)$ & $11.493(47)$ & $11.042(57)$ & $10.620(75)$ & $10.037(46)$ & -1.18 & Class II & BSC & $4300(50)$ & $3.00(25)$ & $67.224(104)$ & 16.0 & 6.19 & $8111_{-3091}^{+1685}$ & 1.02 & 485 & 2.80 \\
58 & G285.4929-00.6920 & 157.62374 & -58.70752 & $20.385(7)$ & $18.826(6)$ & $17.977(7)$ & $17.271(6)$ & $16.950(9)$ & $15.129(16)$ & $13.935(19)$ & $12.723(11)$ & $11.225(51)$ & $10.702(46)$ & $10.332(66)$ & $9.535(25)$ & -0.94 & Class II & BSC & $3600(50)$ & $3.50(25)$ & $0.257(3)$ & 37.4 & 0.52 & $1187_{-321}^{+1493}$ & 1.40 & 497 & 4.04 \\
59 & G286.5450-00.3347 & 159.72849 & -58.92423 & $18.932(7)$ & $17.576(11)$ & $17.101(5)$ & $16.630(5)$ & $16.455(6)$ & $14.971(7)$ & $14.021(10)$ & $13.412(11)$ & $12.160(51)$ & $11.704(68)$ & $11.142(80)$ & $10.469(86)$ & -0.88 & Class II & BSC & $4500(50)$ & $3.00(25)$ & $3.153(9)$ & 16.3 & 1.73 & $3106_{-567}^{+931}$ & 0.98 & 444 & 2.54 \\
60 & G286.8727-00.5898 & 160.04628 & -59.30610 & $19.731(5)$ & $17.954(5)$ & $17.113(5)$ & $16.541(5)$ & $16.314(6)$ & $14.857(7)$ & $13.896(9)$ & $13.285(11)$ & $12.465(56)$ & $12.009(87)$ & $11.534(105)$ & $10.874(82)$ & -1.01 & Class II & BSC & $4300(50)$ & $3.00(25)$ & $2.163(4)$ & 7.7 & 2.40 & $2259_{-322}^{+804}$ & 0.99 & 255 & 0.54 \\
61 & G288.6023-00.1171 & 163.50057 & -59.68111 & $21.854(16)$ & $19.508(8)$ & $18.446(8)$ & $17.502(6)$ & $17.111(6)$ & $15.254(9)$ & $14.044(8)$ & $13.324(11)$ & $12.016(51)$ & $11.648(69)$ & $11.222(64)$ & $10.037(49)$ & -0.59 & Class II & BSC & $4200(50)$ & $3.00(25)$ & $18.632(38)$ & 12.6 & 4.58 & $5949_{-1581}^{+3336}$ & 1.70 & 437 & 3.25 \\
62 & G288.0895-00.2538 & 162.47167 & -59.57641 & $21.396(16)$ & $19.688(10)$ & $18.827(16)$ & $18.091(14)$ & $17.741(18)$ & $16.039(7)$ & $14.870(7)$ & $13.783(11)$ & $12.179(58)$ & $11.672(70)$ & $11.140(65)$ & $10.199(39)$ & -0.57 & Class II & BSC & $3800(50)$ & $3.00(25)$ & $0.938(7)$ & 23.3 & 2.17 & $2592_{-850}^{+952}$ & 1.18 & 471 & 1.33 \\
63 & G282.6875-00.7896 & 153.07559 & -57.26860 & $23.360(65)$ & $20.947(18)$ & $19.647(14)$ & $18.600(9)$ & $18.154(12)$ & $16.453(7)$ & $15.290(11)$ & $14.048(12)$ & $12.146(52)$ & $11.639(77)$ & $10.981(63)$ & $9.708(34)$ & -0.03 & Flat & BSC & $3900(50)$ & $3.00(25)$ & $2.893(14)$ & 18.7 & 4.49 & $3916_{-1304}^{+988}$ & 1.09 & 520 & 0.30 \\
64 & G283.0445-00.3230 & 154.10454 & -57.08417 & $19.087(6)$ & $16.771(5)$ & $15.497(6)$ & $14.596(6)$ & $14.187(6)$ & $12.632(13)$ & $11.906(6)$ & $11.478(12)$ & $10.688(40)$ & $10.476(58)$ & $10.208(56)$ & $10.021(44)$ & -2.07 & Class III & BSC & $5800(50)$ & $3.00(25)$ & $204.989(87)$ & 6.3 & 5.59 & $4337_{-746}^{+648}$ & 0.94 & 490 & 9.94 \\
65 & G283.0273-00.7277 & 153.65943 & -57.40987 & $22.704(46)$ & $20.035(11)$ & $18.511(9)$ & $17.338(8)$ & $16.857(9)$ & $15.038(12)$ & $13.998(8)$ & $13.280(12)$ & $12.209(59)$ & $11.826(90)$ & $11.349(88)$ & $11.260(93)$ & -1.72 & Class III & BS & $3400(50)$ & $3.00(25)$ & $2.298(11)$ & 18.6 & 2.99 & $2674_{-719}^{+730}$ & 1.23 & \dots & \dots \\
66 & G282.3153-00.9046 & 152.39085 & -57.14930 & $22.399(36)$ & $19.831(9)$ & $18.342(8)$ & $17.187(8)$ & $16.763(7)$ & $15.048(7)$ & $14.103(7)$ & $13.363(12)$ & $12.098(50)$ & $11.505(69)$ & $10.996(65)$ & $10.322(39)$ & -0.81 & Class II & BS & $3500(50)$ & $4.00(25)$ & $4.709(10)$ & 9.5 & 4.16 & $3172_{-748}^{+1017}$ & 1.09 & \dots & \dots \\
67 & G283.2628-00.9365 & 153.80378 & -57.71495 & $20.331(9)$ & $18.172(6)$ & $17.151(7)$ & $16.359(7)$ & $15.992(8)$ & $14.371(5)$ & $13.422(6)$ & $13.054(12)$ & $12.306(61)$ & $11.414(61)$ & $10.016(70)$ & $8.865(44)$ & 1.23 & Class I & BSC & $4100(50)$ & $3.00(25)$ & $8.179(11)$ & 2.5 & 3.31 & $3318_{-973}^{+1622}$ & 1.03 & 492 & 13.03 \\
68 & G283.4869-00.4220 & 154.68428 & -57.41151 & $23.105(65)$ & $20.410(12)$ & $19.540(15)$ & $18.518(10)$ & $18.021(12)$ & $15.783(8)$ & $14.603(10)$ & $13.742(12)$ & $12.374(54)$ & $11.935(48)$ & $11.314(81)$ & $10.287(31)$ & -0.42 & Class II & BS & $3300(50)$ & $3.00(25)$ & $2.349(12)$ & 36.8 & 3.09 & $3794_{-757}^{+1109}$ & 0.99 & 473 & 1.46 \\
69 & G283.5407-00.4992 & 154.68894 & -57.50551 & $23.192(85)$ & $20.559(14)$ & $19.008(11)$ & $17.924(8)$ & $17.422(9)$ & $15.462(6)$ & $14.204(12)$ & $13.304(12)$ & $11.855(64)$ & $11.402(80)$ & $10.935(84)$ & $10.185(56)$ & -0.93 & Class II & BS & $3200(50)$ & $3.00(25)$ & $1.973(15)$ & 25.5 & 2.44 & $3139_{-850}^{+583}$ & 1.02 & 9 & -0.19 \\
70 & G284.2242-00.7848 & 155.47202 & -58.11759 & $21.108(15)$ & $19.488(9)$ & $18.577(11)$ & $17.790(7)$ & $17.418(9)$ & $15.793(9)$ & $14.461(9)$ & $13.318(12)$ & $11.814(59)$ & $11.363(75)$ & $10.750(64)$ & $9.679(36)$ & -0.37 & Class II & BSC & $4000(50)$ & $3.00(25)$ & $3.973(25)$ & 23.8 & 2.87 & $4165_{-1068}^{+2185}$ & 1.02 & 521 & 4.22 \\
71 & G283.5165-00.3692 & 154.78408 & -57.38370 & $21.515(14)$ & $19.308(6)$ & $18.326(7)$ & $17.441(6)$ & $17.052(7)$ & $15.388(7)$ & $14.444(8)$ & $13.677(12)$ & $12.312(48)$ & $11.903(88)$ & $11.438(87)$ & $10.668(58)$ & -0.95 & Class II & BS & $3700(50)$ & $3.00(25)$ & $2.222(8)$ & 16.5 & 2.42 & $3251_{-719}^{+975}$ & 0.93 & 152 & 0.58 \\
72 & G283.8748-00.8955 & 154.80409 & -58.02026 & $23.014(48)$ & $20.768(15)$ & $19.542(14)$ & $18.452(7)$ & $17.980(10)$ & $16.186(8)$ & $14.897(13)$ & $13.736(12)$ & $11.941(55)$ & $11.413(72)$ & $10.890(75)$ & $9.385(32)$ & 0.06 & Flat & BS & $3300(50)$ & $3.00(25)$ & $1.082(12)$ & 32.7 & 1.96 & $3118_{-1211}^{+1567}$ & 1.05 & 386 & 3.82 \\
73 & G284.0527-00.8335 & 155.14952 & -58.06557 & $21.224(12)$ & $19.068(5)$ & $17.846(6)$ & $16.804(5)$ & $16.380(6)$ & $14.623(10)$ & $13.512(11)$ & $12.568(12)$ & $11.215(44)$ & $10.626(63)$ & $10.194(57)$ & $9.771(45)$ & -1.20 & Class II & BSC & $4300(50)$ & $3.00(25)$ & $58.865(102)$ & 14.4 & 5.00 & $7225_{-1959}^{+1498}$ & 1.18 & 499 & 0.45 \\
74 & G284.3313-00.9230 & 155.50144 & -58.29179 & $22.680(42)$ & $20.367(13)$ & $19.095(12)$ & $18.059(8)$ & $17.596(10)$ & $15.669(11)$ & $14.422(9)$ & $13.469(12)$ & $12.498(53)$ & $12.032(89)$ & $11.580(102)$ & $11.245(76)$ & -1.39 & Class II & BSC & $3700(50)$ & $3.00(25)$ & $6.399(21)$ & 17.0 & 3.52 & $5335_{-1445}^{+2386}$ & 1.13 & 497 & 1.85 \\
75 & G284.9488-00.7231 & 156.70096 & -58.45152 & $16.593(5)$ & $16.018(4)$ & $15.775(5)$ & $15.481(5)$ & $15.317(5)$ & $14.073(15)$ & $13.213(8)$ & $12.415(12)$ & $11.389(52)$ & $10.968(45)$ & $10.468(66)$ & $9.027(25)$ & -0.15 & Flat & BSC & $6300(50)$ & $3.00(25)$ & $9.197(17)$ & 26.8 & 1.58 & $2894_{-253}^{+346}$ & 1.12 & 520 & 2.23 \\
76 & G285.7083-00.1761 & 158.48153 & -58.37258 & $20.684(11)$ & $19.073(7)$ & $18.409(7)$ & $17.805(6)$ & $17.520(9)$ & $15.652(12)$ & $14.540(14)$ & $13.752(12)$ & $12.924(71)$ & $12.499(96)$ & $12.155(101)$ & $11.386(62)$ & -1.10 & Class II & BSC & $3800(50)$ & $3.00(25)$ & $0.666(4)$ & 27.6 & 0.95 & $2572_{-648}^{+1168}$ & 1.10 & \dots & \dots \\
77 & G285.6646-00.4383 & 158.15622 & -58.57688 & $17.245(5)$ & $16.379(5)$ & $15.959(5)$ & $15.569(5)$ & $15.390(5)$ & $14.228(11)$ & $13.506(11)$ & $12.688(12)$ & $11.373(52)$ & $10.909(57)$ & $10.330(56)$ & $9.402(34)$ & -0.57 & Class II & BS & $9200(100)$ & $3.50(25)$ & $124.195(57)$ & 24.0 & 3.49 & $5342_{-948}^{+1051}$ & 1.08 & 490 & 0.33 \\
78 & G284.8994-00.0691 & 157.26756 & -57.86777 & $20.106(7)$ & $18.236(5)$ & $17.200(6)$ & $16.451(6)$ & $16.126(6)$ & $14.644(9)$ & $13.943(12)$ & $13.435(12)$ & $12.759(54)$ & $12.508(86)$ & $11.982(106)$ & $11.603(91)$ & -1.46 & Class II & BSC & $6800(50)$ & $3.00(25)$ & $127.806(48)$ & 4.9 & 5.35 & $7380_{-2771}^{+1494}$ & 1.30 & \dots & \dots \\
79 & G284.9103-00.6825 & 156.67914 & -58.39672 & $21.172(14)$ & $19.448(9)$ & $18.361(9)$ & $17.500(7)$ & $17.098(8)$ & $15.314(9)$ & $13.967(8)$ & $12.985(12)$ & $11.511(46)$ & $11.041(66)$ & $10.667(68)$ & $10.087(49)$ & -1.23 & Class II & BSC & $4100(50)$ & $3.00(25)$ & $12.342(34)$ & 22.1 & 4.38 & $4722_{-1391}^{+2088}$ & 1.02 & 531 & 3.67 \\
80 & G285.6093-00.0728 & 158.41751 & -58.23353 & $19.306(6)$ & $17.816(7)$ & $17.113(5)$ & $16.581(5)$ & $16.329(5)$ & $14.919(11)$ & $14.091(12)$ & $13.414(12)$ & $12.485(56)$ & $12.044(78)$ & $11.454(99)$ & $10.614(48)$ & -0.67 & Class II & BSC & $4600(50)$ & $3.00(25)$ & $5.401(11)$ & 10.9 & 2.28 & $3681_{-832}^{+1036}$ & 1.12 & 324 & 1.86 \\
81 & G285.6629-00.0846 & 158.49420 & -58.27065 & $18.662(5)$ & $17.465(5)$ & $16.882(5)$ & $16.414(5)$ & $16.195(5)$ & $14.851(11)$ & $13.972(9)$ & $13.185(12)$ & $12.063(46)$ & $11.634(73)$ & $11.187(88)$ & $9.907(38)$ & -0.39 & Class II & BSC & $5300(50)$ & $3.50(25)$ & $10.869(19)$ & 17.9 & 2.41 & $4441_{-878}^{+1020}$ & 0.95 & 484 & 1.78 \\
82 & G285.9756-00.6689 & 158.44935 & -58.93297 & $22.139(24)$ & $20.471(13)$ & $18.654(7)$ & $17.797(6)$ & $17.491(8)$ & $14.874(9)$ & $13.452(10)$ & $12.134(12)$ & $10.448(35)$ & $9.395(49)$ & $8.515(31)$ & $7.637(27)$ & 0.38 & Class I & BSC & $3800(50)$ & $3.00(25)$ & $10.304(39)$ & 35.7 & 5.24 & $3551_{-852}^{+1110}$ & 1.12 & 507 & 7.33 \\
83 & G285.8521-00.0556 & 158.83372 & -58.34011 & $16.365(6)$ & $15.789(6)$ & $15.519(7)$ & $15.214(7)$ & $15.040(7)$ & $13.731(11)$ & $12.790(12)$ & $11.879(12)$ & $10.690(47)$ & $10.350(63)$ & $9.822(45)$ & $8.806(33)$ & -0.66 & Class II & BSC & $6500(50)$ & $3.00(25)$ & $9.604(20)$ & 30.6 & 1.77 & $2409_{-194}^{+232}$ & 1.03 & 486 & 5.91 \\
84 & G285.9064-00.1729 & 158.81203 & -58.46892 & $22.379(26)$ & $19.642(8)$ & $18.230(5)$ & $17.198(5)$ & $16.750(5)$ & $14.867(13)$ & $13.769(10)$ & $13.249(12)$ & $12.531(69)$ & $11.836(52)$ & $10.986(74)$ & $9.383(25)$ & 0.78 & Class I & BS & $3500(50)$ & $3.00(25)$ & $8.742(22)$ & 7.6 & 3.41 & $4762_{-667}^{+708}$ & 0.97 & 477 & 0.76 \\
85 & G286.4939-00.6171 & 159.37343 & -59.14561 & $21.018(11)$ & $18.974(7)$ & $17.916(6)$ & $16.998(5)$ & $16.623(7)$ & $15.045(5)$ & $14.203(10)$ & $13.555(12)$ & $12.338(60)$ & $11.801(80)$ & $11.191(69)$ & $10.373(54)$ & -0.57 & Class II & BSC & $3900(50)$ & $3.00(25)$ & $4.847(13)$ & 8.5 & 2.92 & $3750_{-935}^{+1836}$ & 1.05 & 484 & 0.89 \\
86 & G288.9471-00.2161 & 164.03349 & -59.91958 & $20.195(15)$ & $18.701(12)$ & $17.815(12)$ & $17.153(11)$ & $16.852(13)$ & $15.207(9)$ & $14.080(10)$ & $13.066(12)$ & $11.617(53)$ & $11.299(74)$ & $11.070(82)$ & $10.709(61)$ & -1.82 & Class III & BSC & $3700(50)$ & $3.00(25)$ & $1.492(13)$ & 31.4 & 1.34 & $2460_{-445}^{+496}$ & 0.96 & 428 & 4.21 \\
87 & G288.8386-00.2983 & 163.76724 & -59.94698 & $19.639(7)$ & $17.812(6)$ & $16.865(6)$ & $16.228(6)$ & $15.930(6)$ & $14.219(10)$ & $13.126(6)$ & $12.401(12)$ & $11.581(50)$ & $11.186(62)$ & $10.811(80)$ & $10.176(48)$ & -1.24 & Class II & BS & $3400(50)$ & $3.00(25)$ & $0.045(0)$ & 25.4 & 0.20 & $369_{-70}^{+87}$ & 4.93 & 463 & 2.47 \\
88 & G288.3383-00.7396 & 162.47852 & -60.12220 & $18.348(4)$ & $17.422(5)$ & $17.059(5)$ & $16.684(5)$ & $16.513(6)$ & $14.919(8)$ & $13.632(6)$ & $12.057(12)$ & $9.154(45)$ & $8.218(46)$ & $7.273(34)$ & $6.135(27)$ & 0.64 & Class I & BSC & $5300(50)$ & $4.00(25)$ & $10.643(183)$ & 38.9 & 1.73 & $3790_{-676}^{+949}$ & 1.08 & 490 & 4.12 \\
89 & G283.1762-00.5021 & 154.12245 & -57.30625 & $21.769(16)$ & $19.430(7)$ & $18.175(6)$ & $17.233(6)$ & $16.833(6)$ & $15.169(9)$ & $14.294(8)$ & $13.719(13)$ & $12.875(48)$ & $12.297(69)$ & $11.696(91)$ & $10.889(104)$ & -0.56 & Class II & BSC & $3900(50)$ & $4.50(25)$ & $2.019(3)$ & 3.9 & 3.63 & $2373_{-408}^{+687}$ & 1.12 & 37 & 0.15 \\
90 & G282.5417-00.7996 & 152.84389 & -57.19350 & $22.706(36)$ & $20.553(14)$ & $19.461(12)$ & $18.566(9)$ & $18.185(12)$ & $16.354(7)$ & $15.029(8)$ & $13.957(13)$ & $12.708(72)$ & $12.159(85)$ & $11.429(93)$ & $10.428(71)$ & -0.19 & Flat & BSC & $3700(50)$ & $3.00(25)$ & $0.773(6)$ & 23.0 & 2.56 & $2715_{-757}^{+1111}$ & 1.08 & 525 & 1.36 \\
91 & G284.2723-00.2222 & 156.11627 & -57.66862 & $22.019(28)$ & $19.435(16)$ & $18.052(7)$ & $16.921(11)$ & $16.476(9)$ & $14.620(8)$ & $13.510(10)$ & $12.623(13)$ & $11.234(52)$ & $10.680(59)$ & $10.048(62)$ & $9.419(90)$ & -0.73 & Class II & BS & $3400(50)$ & $3.00(25)$ & $5.602(25)$ & 20.8 & 3.04 & $3332_{-880}^{+2831}$ & 0.95 & 377 & 2.54 \\
92 & G283.8847-00.9616 & 154.75112 & -58.08090 & $22.720(38)$ & $20.278(11)$ & $18.910(9)$ & $17.815(6)$ & $17.344(8)$ & $15.469(9)$ & $14.353(16)$ & $13.546(13)$ & $12.484(47)$ & $11.917(72)$ & $11.462(82)$ & $10.789(77)$ & -0.92 & Class II & BS & $3300(50)$ & $3.00(25)$ & $1.673(7)$ & 19.8 & 2.19 & $3157_{-581}^{+702}$ & 1.00 & 5 & 0.18 \\
93 & G284.1898-00.3326 & 155.87557 & -57.71797 & $22.983(52)$ & $20.298(12)$ & $18.864(9)$ & $17.710(7)$ & $17.237(11)$ & $15.161(7)$ & $13.810(8)$ & $12.900(13)$ & $11.468(51)$ & $11.112(67)$ & $10.831(89)$ & $10.348(73)$ & -1.58 & Class II & BS & $3400(50)$ & $3.00(25)$ & $28.100(124)$ & 27.3 & 3.83 & $8403_{-2994}^{+1587}$ & 1.07 & 379 & 2.32 \\
94 & G284.3311-00.0431 & 156.38671 & -57.54809 & $19.565(6)$ & $17.472(4)$ & $16.309(5)$ & $15.358(5)$ & $14.976(5)$ & $13.446(10)$ & $12.727(10)$ & $12.263(13)$ & $11.805(48)$ & $11.570(58)$ & $11.312(76)$ & $10.940(54)$ & -1.85 & Class III & BSC & $4300(50)$ & $4.50(25)$ & $32.482(34)$ & 4.4 & 3.75 & $4087_{-569}^{+526}$ & 1.01 & 33 & 0.17 \\
95 & G284.9035-00.2253 & 157.12143 & -58.00342 & $17.142(5)$ & $16.407(4)$ & $16.047(5)$ & $15.773(5)$ & $15.625(5)$ & $14.451(10)$ & $13.643(7)$ & $12.823(13)$ & $11.735(60)$ & $11.321(61)$ & $10.806(77)$ & $9.817(42)$ & -0.64 & Class II & BSC & $4600(50)$ & $3.00(25)$ & $1.719(7)$ & 18.3 & 0.57 & $2011_{-159}^{+229}$ & 1.05 & 515 & 1.95 \\
96 & G285.7293-00.7095 & 157.99837 & -58.84331 & $22.098(25)$ & $20.243(12)$ & $18.685(10)$ & $17.751(9)$ & $17.371(10)$ & $15.472(6)$ & $14.204(6)$ & $13.226(13)$ & $12.193(54)$ & $11.715(61)$ & $11.168(71)$ & $10.599(46)$ & -0.99 & Class II & BSC & $3800(50)$ & $3.00(25)$ & $5.419(15)$ & 14.4 & 4.10 & $3909_{-1131}^{+1767}$ & 1.08 & 492 & 1.30 \\
97 & G286.0593-00.7284 & 158.53171 & -59.02644 & $21.289(20)$ & $18.953(8)$ & $18.158(7)$ & $17.357(7)$ & $17.010(9)$ & $15.345(15)$ & $14.326(11)$ & $13.668(13)$ & $12.365(64)$ & $11.772(76)$ & $11.542(82)$ & $10.588(57)$ & -0.89 & Class II & BSC & $6500(50)$ & $3.50(25)$ & $42.490(27)$ & 12.5 & 6.00 & $5231_{-1553}^{+3397}$ & 1.85 & \dots & \dots \\
98 & G285.5440-00.0629 & 158.31973 & -58.19202 & $17.057(10)$ & $16.641(7)$ & $16.583(5)$ & $16.413(5)$ & $16.313(6)$ & $15.323(7)$ & $14.502(9)$ & $13.356(13)$ & $11.146(52)$ & $10.547(64)$ & $9.504(38)$ & $7.488(27)$ & 1.41 & Class I & BSC & $6900(50)$ & $3.00(25)$ & $4.794(20)$ & 28.9 & 1.51 & $2916_{-316}^{+432}$ & 1.05 & 483 & 0.92 \\
99 & G285.8849-00.4171 & 158.54212 & -58.66961 & $20.909(15)$ & $19.510(11)$ & $18.902(12)$ & $18.292(11)$ & $18.008(14)$ & $16.357(11)$ & $15.304(16)$ & $14.266(13)$ & $12.735(63)$ & $12.225(80)$ & $11.879(102)$ & $11.363(67)$ & -1.30 & Class II & BSC & $3900(50)$ & $3.00(25)$ & $0.688(6)$ & 23.5 & 1.49 & $2898_{-929}^{+1501}$ & 1.04 & \dots & \dots \\
100 & G285.7716-00.6377 & 158.13917 & -58.80295 & $21.279(15)$ & $18.748(7)$ & $17.317(5)$ & $16.396(5)$ & $15.990(6)$ & $14.273(11)$ & $13.404(9)$ & $12.721(13)$ & $12.368(58)$ & $12.022(67)$ & $11.539(74)$ & $11.447(77)$ & -1.74 & Class III & BSC & $5600(50)$ & $3.00(25)$ & $55.586(24)$ & 1.6 & 6.22 & $4423_{-691}^{+1267}$ & 1.02 & 481 & 4.58 \\
101 & G288.0752-00.3988 & 162.31685 & -59.69926 & $21.865(21)$ & $19.397(8)$ & $18.031(7)$ & $17.004(5)$ & $16.586(6)$ & $14.914(6)$ & $14.101(7)$ & $13.579(13)$ & $12.225(55)$ & $11.820(62)$ & $11.579(85)$ & $11.269(82)$ & -1.77 & Class III & BS & $3900(50)$ & $4.50(25)$ & $28.267(45)$ & 9.2 & 4.42 & $6883_{-3143}^{+2147}$ & 1.05 & 470 & 2.16 \\
102 & G282.4980-00.3827 & 153.21438 & -56.82607 & $22.528(31)$ & $20.041(10)$ & $18.722(8)$ & $17.723(7)$ & $17.328(8)$ & $15.723(6)$ & $14.840(7)$ & $14.128(14)$ & $12.899(64)$ & $12.089(81)$ & $11.589(89)$ & $10.854(50)$ & -0.54 & Class II & KU03 & $4500(125)$ & $5.00(25)$ & $4.109(7)$ & 10.2 & 4.57 & $3540_{-769}^{+804}$ & 1.00 & 392 & 0.41 \\
103 & G283.2980-00.4850 & 154.32751 & -57.35974 & $21.697(22)$ & $19.645(10)$ & $18.578(10)$ & $17.718(9)$ & $17.342(10)$ & $15.646(9)$ & $14.723(11)$ & $13.931(14)$ & $12.111(46)$ & $11.684(71)$ & $11.415(87)$ & $10.418(51)$ & -0.95 & Class II & BSC & $3700(50)$ & $3.50(25)$ & $0.854(4)$ & 16.5 & 2.30 & $2261_{-846}^{+571}$ & 0.98 & 475 & 1.58 \\
104 & G284.2961-00.7966 & 155.57465 & -58.16640 & $19.572(6)$ & $18.175(5)$ & $17.472(7)$ & $16.926(6)$ & $16.675(7)$ & $15.306(9)$ & $14.377(9)$ & $13.579(14)$ & $12.282(59)$ & $11.974(84)$ & $11.613(101)$ & $10.923(61)$ & -1.28 & Class II & BSC & $4200(50)$ & $3.00(25)$ & $2.823(15)$ & 23.6 & 1.66 & $3405_{-631}^{+1147}$ & 0.99 & 518 & 1.73 \\
105 & G284.5841-00.4582 & 156.37742 & -58.03414 & $21.279(13)$ & $18.774(6)$ & $17.282(6)$ & $16.257(5)$ & $15.799(5)$ & $14.005(9)$ & $13.049(12)$ & $12.335(14)$ & $11.202(42)$ & $10.984(64)$ & $10.640(65)$ & $10.453(59)$ & -1.95 & Class III & BS & $3200(50)$ & $4.00(25)$ & $19.124(64)$ & 15.8 & 2.89 & $4918_{-970}^{+834}$ & 1.07 & 520 & 3.23 \\
106 & G284.0757-00.8929 & 155.12479 & -58.12789 & $18.088(4)$ & $16.713(4)$ & $16.147(5)$ & $15.662(5)$ & $15.429(5)$ & $14.004(9)$ & $13.132(8)$ & $12.457(14)$ & $11.202(44)$ & $10.742(59)$ & $10.249(59)$ & $9.107(45)$ & -0.45 & Class II & BSC & $4400(50)$ & $3.00(25)$ & $4.439(13)$ & 16.3 & 1.50 & $2527_{-238}^{+348}$ & 1.05 & 52 & 2.33 \\
107 & G285.0051-00.8359 & 156.67911 & -58.57712 & $17.646(5)$ & $16.922(4)$ & $16.563(5)$ & $16.247(5)$ & $16.065(6)$ & $14.882(11)$ & $14.113(8)$ & $13.379(14)$ & $11.990(65)$ & $11.743(72)$ & $10.996(77)$ & $9.774(30)$ & -0.23 & Flat & KU03 & $8000(125)$ & $3.50(25)$ & $32.375(40)$ & 35.7 & 2.82 & $4890_{-1182}^{+2009}$ & 1.68 & 516 & 1.82 \\
108 & G285.3252-00.7926 & 157.24730 & -58.70714 & $22.528(38)$ & $20.482(17)$ & $19.343(15)$ & $18.498(11)$ & $18.115(15)$ & $16.328(10)$ & $15.071(12)$ & $14.020(14)$ & $12.741(43)$ & $12.237(90)$ & $11.964(73)$ & $10.975(65)$ & -0.88 & Class II & BSC & $4300(50)$ & $3.50(25)$ & $7.682(12)$ & 19.3 & 5.16 & $5105_{-3198}^{+1552}$ & 1.09 & 499 & 0.80 \\
109 & G285.9851-00.5168 & 158.61283 & -58.80618 & $20.803(12)$ & $18.867(6)$ & $17.836(6)$ & $17.201(6)$ & $16.937(7)$ & $15.279(12)$ & $14.256(13)$ & $13.544(14)$ & $12.724(71)$ & $12.086(84)$ & $11.681(102)$ & $10.583(55)$ & -0.44 & Class II & BSC & $4200(50)$ & $3.00(25)$ & $2.164(6)$ & 9.0 & 3.01 & $2604_{-585}^{+691}$ & 1.04 & 305 & 3.74 \\
110 & G285.4763-00.2024 & 158.07475 & -58.27805 & $19.569(6)$ & $18.111(6)$ & $17.359(6)$ & $16.783(6)$ & $16.500(7)$ & $15.140(8)$ & $14.394(12)$ & $13.714(14)$ & $12.340(61)$ & $11.581(60)$ & $11.164(70)$ & $10.475(41)$ & -0.76 & Class II & BSC & $4300(50)$ & $3.00(25)$ & $2.946(10)$ & 11.8 & 1.99 & $3210_{-734}^{+770}$ & 1.01 & 7 & 0.28 \\
\hline
\end{tabular}
\end{sidewaystable*}

\addtocounter{table}{-1}
\begin{sidewaystable*}
\caption{Photometric, astrometric, and physical properties of the MIR-excess candidates (continued).}
\centering
\tiny
\setlength{\tabcolsep}{0.8pt}
\renewcommand{\arraystretch}{0.92}
\begin{tabular}{c l c c c c c c c c c c c c c c c c c c c c c c c c c c}
\hline\hline
\# & GLIMPSE ID & RA & Dec & $g$ & $r$ & $i$ & $z$ & $Y$ & $J$ & $H$ & $K_s$ & $[3.6]$ & $[4.5]$ & $[5.8]$ & $[8.0]$ & $\alpha_{\rm IRAC}$ & Class & Model & $T_{\rm eff}$ & $\log g$ & $L_{\rm bol}$ & $V_{\rm gfb}$ & $A_V$ & $D_{\rm photogeo}$ & RUWE & $n_{\rm pairs}$ & $J_{\rm W1W2}$ \\
 &  & (deg) & (deg) & (mag) & (mag) & (mag) & (mag) & (mag) & (mag) & (mag) & (mag) & (mag) & (mag) & (mag) & (mag) &  &  &  & (K) & (dex) & ($L_\odot$) &  & (mag) & (pc) &  &  &  \\
\hline
111 & G286.3903-00.6162 & 159.19876 & -59.09377 & $23.291(77)$ & $20.823(18)$ & $19.527(14)$ & $18.537(10)$ & $18.091(16)$ & $16.114(8)$ & $14.691(10)$ & $13.725(14)$ & $12.278(59)$ & $11.596(60)$ & $11.077(78)$ & $10.491(34)$ & -0.81 & Class II & BSC & $3500(50)$ & $3.00(25)$ & $4.158(23)$ & 25.1 & 3.59 & $4936_{-1255}^{+1621}$ & 1.09 & 492 & 1.67 \\
112 & G283.8519-00.6809 & 154.98927 & -57.82811 & $23.271(65)$ & $20.657(13)$ & $19.358(11)$ & $18.225(7)$ & $17.720(9)$ & $15.843(8)$ & $14.534(10)$ & $13.422(15)$ & $11.810(54)$ & $11.186(86)$ & $10.587(63)$ & $9.610(35)$ & -0.32 & Class II & BS & $3100(50)$ & $3.00(25)$ & $1.077(11)$ & 33.6 & 1.73 & $2844_{-1273}^{+1154}$ & 1.47 & 496 & 0.65 \\
113 & G283.8917-00.8988 & 154.82736 & -58.03227 & $22.306(25)$ & $20.063(9)$ & $18.772(8)$ & $17.699(6)$ & $17.259(7)$ & $15.531(12)$ & $14.617(16)$ & $13.887(15)$ & $12.686(59)$ & $12.123(71)$ & $11.583(105)$ & $10.589(56)$ & -0.45 & Class II & BS & $3500(50)$ & $3.00(25)$ & $1.904(8)$ & 15.0 & 2.65 & $3309_{-696}^{+1143}$ & 1.02 & 163 & 0.49 \\
114 & G283.9174-00.9231 & 154.84269 & -58.06669 & $22.794(43)$ & $20.470(16)$ & $19.206(10)$ & $18.180(7)$ & $17.729(9)$ & $16.013(7)$ & $14.993(18)$ & $13.970(15)$ & $12.238(49)$ & $11.578(75)$ & $10.887(70)$ & $9.730(36)$ & 0.04 & Flat & BS & $3300(50)$ & $3.00(25)$ & $1.293(9)$ & 25.7 & 2.02 & $3346_{-927}^{+813}$ & 0.99 & 325 & 0.08 \\
115 & G284.4746-00.6760 & 155.98280 & -58.16039 & $18.410(6)$ & $17.101(4)$ & $16.409(5)$ & $15.856(5)$ & $15.603(5)$ & $14.316(12)$ & $13.540(12)$ & $12.765(15)$ & $11.563(45)$ & $11.046(63)$ & $10.255(56)$ & $9.012(38)$ & 0.13 & Flat & BSC & $5800(50)$ & $3.00(25)$ & $22.022(26)$ & 15.2 & 3.19 & $4090_{-776}^{+1105}$ & 1.07 & 518 & 0.42 \\
116 & G284.8286-00.5315 & 156.69768 & -58.22535 & $22.083(27)$ & $19.561(9)$ & $18.100(8)$ & $17.031(6)$ & $16.555(7)$ & $14.763(11)$ & $13.797(15)$ & $13.068(15)$ & $12.500(67)$ & $12.101(79)$ & $11.718(105)$ & $11.033(74)$ & -1.17 & Class II & BSC & $4400(50)$ & $3.00(25)$ & $52.793(62)$ & 5.0 & 5.59 & $7014_{-1540}^{+1384}$ & 1.02 & 511 & 1.83 \\
117 & G285.0185-00.3072 & 157.22690 & -58.13311 & $21.992(28)$ & $19.412(8)$ & $17.993(9)$ & $16.876(6)$ & $16.416(7)$ & $14.564(15)$ & $13.602(8)$ & $12.911(15)$ & $12.124(56)$ & $11.712(74)$ & $11.295(79)$ & $11.014(66)$ & -1.56 & Class II & BS & $3400(50)$ & $4.00(25)$ & $17.598(34)$ & 6.3 & 3.82 & $5428_{-1508}^{+919}$ & 1.11 & 505 & 0.78 \\
118 & G284.9429-00.8957 & 156.51738 & -58.59519 & $18.214(6)$ & $17.075(5)$ & $16.736(6)$ & $16.403(6)$ & $16.129(7)$ & $14.598(12)$ & $13.841(16)$ & $13.250(15)$ & $12.301(58)$ & $11.782(67)$ & $11.268(85)$ & $10.454(50)$ & -0.73 & Class II & BSC & $5100(50)$ & $3.50(25)$ & $3.572(9)$ & 20.5 & 1.64 & $2901_{-380}^{+424}$ & 0.98 & \dots & \dots \\
119 & G285.1809-00.8432 & 156.95949 & -58.67530 & $18.516(6)$ & $17.379(6)$ & $16.828(7)$ & $16.277(7)$ & $16.014(8)$ & $14.549(12)$ & $13.568(13)$ & $12.616(15)$ & $11.355(34)$ & $10.892(39)$ & $10.292(35)$ & $9.026(24)$ & -0.16 & Flat & BSC & $5700(50)$ & $3.50(25)$ & $14.548(26)$ & 27.0 & 2.74 & $4233_{-904}^{+1456}$ & 1.06 & 501 & 2.93 \\
120 & G286.3763-00.7749 & 159.02194 & -59.22482 & $22.786(45)$ & $20.257(12)$ & $19.020(9)$ & $18.074(7)$ & $17.628(9)$ & $15.876(19)$ & $14.674(19)$ & $13.735(15)$ & $12.294(52)$ & $11.864(71)$ & $11.429(89)$ & $10.460(51)$ & -0.75 & Class II & BSC & $4700(50)$ & $3.50(25)$ & $49.662(73)$ & 13.8 & 5.98 & $9186_{-2783}^{+5571}$ & 1.43 & 497 & 2.00 \\
121 & G286.9290-00.9956 & 159.75602 & -59.68840 & $20.526(12)$ & $19.000(9)$ & $18.153(9)$ & $17.460(8)$ & $17.174(15)$ & $15.747(7)$ & $14.971(10)$ & $14.166(15)$ & $13.041(69)$ & $12.390(103)$ & $12.141(104)$ & $11.466(79)$ & -1.11 & Class II & BSC & $6200(50)$ & $3.00(25)$ & $71.805(69)$ & 12.5 & 4.45 & $11044_{-5694}^{+3255}$ & 1.02 & 329 & 0.20 \\
122 & G288.4956-00.6287 & 162.86001 & -60.09336 & $18.717(6)$ & $16.799(6)$ & $15.875(5)$ & $15.202(5)$ & $14.899(5)$ & $13.423(18)$ & $12.576(21)$ & $12.211(15)$ & $11.319(66)$ & $11.390(73)$ & $11.236(83)$ & $11.191(77)$ & -2.66 & Class III & \dots & \dots & \dots & \dots & \dots & \dots & \dots & 1.10 & 475 & 0.40 \\
123 & G283.3465-00.4417 & 154.44698 & -57.35049 & $22.765(41)$ & $20.784(16)$ & $19.687(15)$ & $18.795(10)$ & $18.389(14)$ & $16.641(12)$ & $15.337(13)$ & $14.062(16)$ & $12.524(53)$ & $11.975(61)$ & $11.416(105)$ & $10.373(50)$ & -0.38 & Class II & BSC & $4200(50)$ & $3.00(25)$ & $5.486(19)$ & 23.5 & 4.77 & $5281_{-736}^{+1411}$ & 0.95 & 487 & 0.60 \\
124 & G284.1397-00.4567 & 155.67137 & -57.79577 & $21.244(14)$ & $19.314(8)$ & $18.247(9)$ & $17.380(7)$ & $17.018(8)$ & $15.456(7)$ & $14.337(10)$ & $13.212(16)$ & $11.557(53)$ & $10.932(62)$ & $10.181(64)$ & $8.770(32)$ & 0.37 & Class I & BSC & $5200(50)$ & $3.50(25)$ & $25.556(55)$ & 19.6 & 5.08 & $5759_{-1751}^{+3961}$ & 1.02 & 505 & 1.50 \\
125 & G284.2278-00.7400 & 155.52358 & -58.08194 & $22.321(28)$ & $20.647(15)$ & $19.474(14)$ & $18.607(10)$ & $18.234(13)$ & $16.496(8)$ & $15.304(10)$ & $14.279(16)$ & $12.680(71)$ & $12.350(81)$ & $11.507(91)$ & $10.542(52)$ & -0.30 & Flat & BSC & $4000(50)$ & $3.00(25)$ & $4.994(31)$ & 27.7 & 3.88 & $5602_{-2310}^{+2727}$ & 1.07 & 451 & 1.54 \\
126 & G284.2701-00.8161 & 155.51314 & -58.16881 & $19.378(6)$ & $18.272(6)$ & $17.710(7)$ & $17.158(6)$ & $16.907(8)$ & $15.371(8)$ & $14.249(10)$ & $13.211(16)$ & $11.986(50)$ & $11.419(60)$ & $11.097(59)$ & $9.835(41)$ & -0.45 & Class II & BSC & $5700(50)$ & $3.50(25)$ & $7.578(20)$ & 29.7 & 2.92 & $4250_{-1103}^{+1409}$ & 0.91 & 516 & 1.32 \\
127 & G284.6586-00.7789 & 156.17443 & -58.34552 & $21.332(15)$ & $18.953(8)$ & $17.688(8)$ & $16.749(8)$ & $16.312(8)$ & $14.537(9)$ & $13.286(21)$ & $12.382(16)$ & $10.977(48)$ & $10.559(65)$ & $10.126(51)$ & $9.588(41)$ & -1.24 & Class II & BSC & $4700(50)$ & $3.00(25)$ & $66.470(114)$ & 14.5 & 5.53 & $6450_{-2310}^{+1592}$ & 1.02 & 471 & 4.39 \\
128 & G284.1854-00.6071 & 155.59151 & -57.94713 & $22.647(45)$ & $20.196(12)$ & $19.034(13)$ & $18.197(9)$ & $17.796(11)$ & $16.035(8)$ & $14.843(8)$ & $14.069(16)$ & $12.989(73)$ & $12.402(77)$ & $11.714(91)$ & $11.125(83)$ & -0.67 & Class II & BSC & $4500(50)$ & $3.50(25)$ & $12.182(22)$ & 9.7 & 5.12 & $5940_{-1867}^{+1789}$ & 1.07 & 512 & 0.65 \\
129 & G284.1996-00.9232 & 155.29077 & -58.22066 & $21.360(20)$ & $19.702(10)$ & $19.053(14)$ & $18.223(10)$ & $17.849(13)$ & $15.972(7)$ & $14.790(7)$ & $13.763(16)$ & $12.373(62)$ & $11.852(66)$ & $11.411(88)$ & $10.587(63)$ & -0.81 & Class II & BSC & $4000(50)$ & $3.00(25)$ & $3.246(17)$ & 25.7 & 3.10 & $4232_{-1273}^{+1173}$ & 1.03 & 522 & 1.51 \\
130 & G284.3993-00.8119 & 155.72403 & -58.23483 & $20.706(11)$ & $18.591(6)$ & $17.482(8)$ & $16.484(7)$ & $16.056(8)$ & $14.430(11)$ & $13.657(10)$ & $13.166(16)$ & $11.880(49)$ & $11.595(60)$ & $11.270(101)$ & $11.188(68)$ & -2.03 & Class III & BSC & $4500(50)$ & $3.00(25)$ & $33.758(54)$ & 13.8 & 4.27 & $5954_{-1280}^{+966}$ & 0.96 & 526 & 1.74 \\
131 & G285.2448-00.6232 & 157.28363 & -58.52064 & $19.846(9)$ & $18.622(7)$ & $17.983(7)$ & $17.464(7)$ & $17.216(11)$ & $15.863(10)$ & $14.980(14)$ & $14.160(16)$ & $12.936(41)$ & $12.445(64)$ & $12.102(85)$ & $10.992(42)$ & -0.66 & Class II & BSC & $5200(50)$ & $3.50(25)$ & $3.961(6)$ & 17.1 & 2.76 & $4015_{-873}^{+1242}$ & 1.02 & 496 & 0.76 \\
132 & G286.0859-00.6231 & 158.67872 & -58.94865 & $19.640(6)$ & $17.946(5)$ & $17.302(6)$ & $16.632(6)$ & $16.350(8)$ & $14.322(12)$ & $13.408(10)$ & $12.597(16)$ & $11.623(56)$ & $11.163(63)$ & $10.734(63)$ & $9.783(43)$ & -0.75 & Class II & BSC & $5500(50)$ & $3.50(25)$ & $14.369(26)$ & 26.8 & 3.79 & $3839_{-876}^{+2210}$ & 0.97 & 492 & 2.50 \\
133 & G285.9058-00.9600 & 158.04613 & -59.14878 & $21.464(15)$ & $19.812(9)$ & $18.868(8)$ & $18.043(7)$ & $17.669(9)$ & $15.867(10)$ & $14.544(10)$ & $13.415(16)$ & $11.960(41)$ & $11.420(73)$ & $11.005(63)$ & $9.859(41)$ & -0.47 & Class II & BSC & $4200(50)$ & $3.00(25)$ & $6.492(21)$ & 25.9 & 3.96 & $4732_{-1086}^{+1565}$ & 1.07 & 500 & 2.05 \\
134 & G282.0144-00.0541 & 152.83350 & -56.28053 & $22.613(28)$ & $19.609(7)$ & $17.806(5)$ & $16.599(5)$ & $16.101(5)$ & $14.113(8)$ & $12.927(12)$ & $12.412(17)$ & $11.342(51)$ & $11.314(82)$ & $11.361(92)$ & $11.330(95)$ & -2.85 & Class III & BS & $3600(50)$ & $3.50(25)$ & $19.396(13)$ & 2.0 & 5.34 & $3695_{-739}^{+647}$ & 1.12 & 510 & 0.44 \\
135 & G283.2985-00.5876 & 154.22270 & -57.44535 & $22.716(38)$ & $20.469(13)$ & $19.217(13)$ & $18.249(10)$ & $17.814(12)$ & $16.082(15)$ & $14.955(11)$ & $14.064(17)$ & $12.548(47)$ & $11.972(77)$ & $11.415(92)$ & $10.359(74)$ & -0.34 & Class II & BSC & $3900(50)$ & $3.00(25)$ & $2.335(7)$ & 13.3 & 3.99 & $3412_{-1486}^{+1482}$ & 1.15 & 479 & 1.27 \\
136 & G283.4208-00.8449 & 154.14523 & -57.72721 & $23.523(70)$ & $20.532(12)$ & $18.870(8)$ & $17.568(6)$ & $17.013(6)$ & $14.956(8)$ & $13.792(11)$ & $13.016(17)$ & $12.154(36)$ & $11.663(80)$ & $11.302(62)$ & $11.114(74)$ & -1.66 & Class III & BS & $3000(50)$ & $4.00(25)$ & $8.091(12)$ & 13.4 & 4.68 & $3898_{-877}^{+1563}$ & 1.04 & 489 & 2.68 \\
137 & G284.6632-00.1866 & 156.77362 & -57.84493 & $19.791(7)$ & $18.199(5)$ & $17.438(6)$ & $16.927(6)$ & $16.671(6)$ & $15.228(13)$ & $14.287(17)$ & $13.574(17)$ & $12.170(46)$ & $11.577(79)$ & $11.091(76)$ & $10.577(69)$ & -1.02 & Class II & BSC & $3900(50)$ & $3.00(25)$ & $1.171(5)$ & 14.6 & 1.18 & $2417_{-418}^{+712}$ & 1.06 & 519 & 1.09 \\
138 & G284.2793-00.6084 & 155.73939 & -57.99877 & $16.464(6)$ & $15.568(4)$ & $15.389(5)$ & $15.106(5)$ & $14.941(5)$ & $13.846(8)$ & $13.052(13)$ & $12.263(17)$ & $11.275(49)$ & $10.905(66)$ & $10.470(56)$ & $9.322(46)$ & -0.61 & Class II & BSC & $5000(50)$ & $3.50(25)$ & $3.780(10)$ & 18.0 & 0.80 & $2081_{-90}^{+109}$ & 1.11 & 520 & 3.05 \\
139 & G284.9938-00.5270 & 156.96980 & -58.30801 & $21.391(14)$ & $19.466(7)$ & $18.435(8)$ & $17.741(6)$ & $17.446(8)$ & $16.014(16)$ & $14.813(16)$ & $13.850(17)$ & $12.661(102)$ & $12.086(77)$ & $11.744(68)$ & $11.055(56)$ & -1.05 & Class II & BSC & $4000(50)$ & $3.00(25)$ & $3.214(23)$ & 15.3 & 2.48 & $4479_{-1040}^{+1493}$ & 1.17 & \dots & \dots \\
140 & G285.6433-00.9965 & 157.57066 & -59.04601 & $19.234(7)$ & $17.965(6)$ & $17.136(6)$ & $16.681(6)$ & $16.502(8)$ & $15.092(7)$ & $13.881(9)$ & $12.541(17)$ & $9.707(39)$ & $8.726(49)$ & $7.403(28)$ & $5.937(25)$ & 1.57 & Class I & BSC & $4000(50)$ & $3.00(25)$ & $2.586(38)$ & 26.3 & 1.22 & $2390_{-662}^{+2003}$ & 1.61 & 501 & 16.25 \\
141 & G285.8758-00.9895 & 157.96652 & -59.15900 & $21.142(17)$ & $19.259(10)$ & $18.350(9)$ & $17.554(8)$ & $17.212(10)$ & $15.393(8)$ & $14.285(10)$ & $13.502(17)$ & $12.649(65)$ & $11.859(73)$ & $11.209(69)$ & $10.336(61)$ & -0.20 & Flat & BSC & $4600(50)$ & $3.00(25)$ & $13.382(20)$ & 13.4 & 4.60 & $4980_{-1521}^{+7653}$ & 1.01 & 504 & 1.82 \\
142 & G288.7093-00.8769 & 163.02414 & -60.41061 & $15.379(4)$ & $14.938(5)$ & $14.717(4)$ & $14.479(4)$ & $14.353(5)$ & $13.214(7)$ & $12.447(11)$ & $11.709(17)$ & $10.350(45)$ & $10.035(59)$ & $9.640(46)$ & $8.711(28)$ & -0.97 & Class II & BSC & $6700(50)$ & $3.00(25)$ & $24.463(45)$ & 27.0 & 1.31 & $3138_{-683}^{+1472}$ & 5.98 & 475 & 5.08 \\
143 & G283.2175-00.7526 & 153.92634 & -57.53708 & $21.509(19)$ & $18.896(7)$ & $16.811(6)$ & $15.614(6)$ & $15.095(6)$ & $13.141(17)$ & $11.963(21)$ & $11.568(18)$ & $10.865(48)$ & $10.828(55)$ & $10.856(87)$ & $10.706(66)$ & -2.69 & Class III & BS & $3300(50)$ & $3.50(25)$ & $18.638(51)$ & 5.4 & 3.53 & $3162_{-722}^{+570}$ & 1.41 & 504 & 0.25 \\
144 & G284.6808-00.3042 & 156.68578 & -57.95430 & $22.626(35)$ & $19.563(7)$ & $17.905(6)$ & $16.614(5)$ & $16.076(5)$ & $14.156(16)$ & $13.104(16)$ & $12.560(18)$ & $12.041(48)$ & $11.783(76)$ & $11.539(91)$ & $11.189(101)$ & -1.87 & Class III & BS & $3300(50)$ & $4.50(25)$ & $12.714(32)$ & 8.4 & 3.81 & $4077_{-580}^{+705}$ & 1.09 & 169 & 0.55 \\
145 & G286.0815-00.3193 & 158.96375 & -58.68306 & $19.550(5)$ & $17.463(5)$ & $16.309(4)$ & $15.464(4)$ & $15.067(5)$ & $13.691(13)$ & $12.971(12)$ & $12.480(18)$ & $11.484(25)$ & $11.191(67)$ & $10.908(42)$ & $10.569(102)$ & -1.80 & Class III & BSC & $4800(50)$ & $3.00(25)$ & $71.933(62)$ & 8.5 & 4.26 & $5531_{-858}^{+2445}$ & 1.00 & 489 & 4.56 \\
146 & G286.1284-00.3641 & 158.99931 & -58.74516 & $22.826(47)$ & $20.301(13)$ & $18.957(8)$ & $17.957(6)$ & $17.554(9)$ & $15.808(9)$ & $14.855(10)$ & $14.158(18)$ & $12.999(73)$ & $12.354(99)$ & $11.479(85)$ & $10.274(59)$ & 0.33 & Class I & BS & $3700(50)$ & $4.50(25)$ & $3.921(10)$ & 8.9 & 4.23 & $3984_{-907}^{+1160}$ & 1.03 & 44 & 0.74 \\
147 & G284.4742-00.6972 & 155.96073 & -58.17812 & $19.918(10)$ & $18.240(8)$ & $17.369(11)$ & $16.558(10)$ & $16.199(11)$ & $14.219(11)$ & $13.157(14)$ & $12.209(19)$ & $11.154(51)$ & $10.736(66)$ & $10.351(43)$ & $9.605(32)$ & -1.08 & Class II & BSC & $3900(50)$ & $3.00(25)$ & $6.700(28)$ & 22.7 & 2.55 & $3299_{-664}^{+780}$ & 1.14 & 526 & 2.89 \\
148 & G285.9862-00.1481 & 158.96803 & -58.48710 & $22.177(24)$ & $20.160(11)$ & $18.737(8)$ & $17.859(6)$ & $17.504(7)$ & $15.728(12)$ & $14.656(13)$ & $14.037(19)$ & $13.086(43)$ & $12.704(75)$ & $12.110(97)$ & $11.752(68)$ & -1.26 & Class II & BSC & $4000(50)$ & $3.00(25)$ & $5.656(8)$ & 6.6 & 4.30 & $4349_{-925}^{+1748}$ & 0.99 & 55 & 0.72 \\
149 & G284.9366-00.9611 & 156.44069 & -58.64740 & $19.643(7)$ & $18.136(5)$ & $17.416(6)$ & $16.880(5)$ & $16.625(6)$ & $15.067(10)$ & $14.050(12)$ & $13.175(19)$ & $11.903(63)$ & $11.292(77)$ & $10.723(69)$ & $9.812(35)$ & -0.45 & Class II & BSC & $4200(50)$ & $3.00(25)$ & $2.232(11)$ & 19.0 & 1.90 & $2751_{-476}^{+606}$ & 0.99 & 508 & 1.06 \\
150 & G285.3956-00.7030 & 157.45229 & -58.66689 & $21.621(14)$ & $19.700(8)$ & $18.727(8)$ & $17.963(7)$ & $17.635(9)$ & $16.060(12)$ & $15.015(13)$ & $14.137(19)$ & $12.865(40)$ & $12.320(79)$ & $11.974(89)$ & $10.957(64)$ & -0.71 & Class II & BSC & $4500(50)$ & $3.00(25)$ & $3.687(7)$ & 13.7 & 3.80 & $4015_{-1436}^{+1443}$ & 0.99 & 504 & 1.27 \\
151 & G288.5174-00.2322 & 163.24897 & -59.74745 & $20.172(7)$ & $17.934(6)$ & $16.833(5)$ & $16.063(5)$ & $15.721(5)$ & $14.111(21)$ & $13.186(19)$ & $12.818(19)$ & $11.682(80)$ & $11.505(77)$ & $11.601(100)$ & $11.554(89)$ & -2.76 & Class III & \dots & \dots & \dots & \dots & \dots & \dots & \dots & 1.06 & 467 & 0.76 \\
152 & G284.8216-00.2428 & 156.97240 & -57.97574 & $23.458(74)$ & $20.247(11)$ & $18.442(7)$ & $17.011(6)$ & $16.400(6)$ & $14.017(14)$ & $12.622(28)$ & $11.886(20)$ & $11.226(44)$ & $11.056(58)$ & $10.880(68)$ & $10.691(55)$ & -2.23 & Class III & BS & $3200(50)$ & $3.00(25)$ & $32.597(134)$ & 32.9 & 3.93 & $6022_{-1535}^{+901}$ & 0.99 & 514 & 2.02 \\
153 & G283.1471-00.8956 & 153.66813 & -57.61604 & $23.205(57)$ & $21.210(21)$ & $20.064(23)$ & $19.072(15)$ & $18.667(20)$ & $17.102(12)$ & $15.993(18)$ & $15.101(21)$ & $13.769(90)$ & $13.149(94)$ & $11.251(97)$ & $8.914(23)$ & 2.96 & Class I & BSC & $4600(50)$ & $3.50(25)$ & $2.233(6)$ & 14.2 & 5.00 & $4048_{-671}^{+1872}$ & 1.02 & \dots & \dots \\
154 & G288.9987-00.8945 & 163.53641 & -60.55368 & $16.998(5)$ & $15.708(6)$ & $15.173(5)$ & $14.803(5)$ & $14.634(5)$ & $13.525(15)$ & $12.792(20)$ & $12.546(21)$ & $11.463(50)$ & $11.557(78)$ & $11.417(91)$ & $11.598(78)$ & -2.95 & Class III & \dots & \dots & \dots & \dots & \dots & \dots & \dots & 1.04 & 451 & 0.46 \\
155 & G282.7184-00.0477 & 153.89085 & -56.67363 & $19.477(8)$ & $17.081(6)$ & $15.803(7)$ & $14.817(7)$ & $14.431(7)$ & $12.964(12)$ & $12.144(5)$ & $11.646(22)$ & $10.920(44)$ & $10.560(50)$ & $10.227(46)$ & $10.030(42)$ & -1.82 & Class III & BSC & $5900(50)$ & $3.00(25)$ & $133.383(46)$ & 2.4 & 5.77 & $3799_{-666}^{+1121}$ & 1.05 & 510 & 2.67 \\
156 & G284.7940-00.2868 & 156.88473 & -57.99877 & $19.134(7)$ & $17.937(5)$ & $17.204(6)$ & $16.677(5)$ & $16.415(6)$ & $15.083(14)$ & $14.090(19)$ & $13.267(22)$ & $11.710(47)$ & $11.268(41)$ & $10.760(65)$ & $9.980(34)$ & -0.85 & Class II & BSC & $4500(50)$ & $3.00(25)$ & $3.196(13)$ & 20.4 & 1.91 & $3107_{-703}^{+1003}$ & 1.01 & 510 & 2.85 \\
157 & G286.1030-00.2619 & 159.05434 & -58.64379 & $23.079(53)$ & $20.760(15)$ & $19.056(7)$ & $17.946(5)$ & $17.443(6)$ & $15.278(16)$ & $13.951(14)$ & $12.865(22)$ & $11.504(47)$ & $10.988(61)$ & $10.558(82)$ & $9.757(39)$ & -0.86 & Class II & BSC & $3800(50)$ & $3.00(25)$ & $9.259(27)$ & 19.9 & 5.44 & $4036_{-1156}^{+1545}$ & 1.14 & 493 & 2.27 \\
158 & G286.1014-00.1207 & 159.18547 & -58.52033 & $18.454(5)$ & $16.926(10)$ & $16.169(4)$ & $15.749(5)$ & $15.580(5)$ & $14.267(16)$ & $13.498(13)$ & $12.913(22)$ & $11.716(45)$ & $11.155(78)$ & $10.337(72)$ & $9.513(91)$ & -0.26 & Flat & BS & $3900(50)$ & $3.00(25)$ & $2.945(10)$ & 11.3 & 0.89 & $2558_{-289}^{+409}$ & 1.01 & 491 & 2.00 \\
159 & G285.5585-00.8778 & 157.54808 & -58.90057 & $21.056(12)$ & $19.517(8)$ & $18.747(8)$ & $18.059(7)$ & $17.772(10)$ & $16.379(11)$ & $15.591(14)$ & $14.928(22)$ & $13.214(66)$ & $12.554(103)$ & $11.996(84)$ & $10.352(38)$ & 0.39 & Class I & BSC & $4700(50)$ & $3.00(25)$ & $1.183(4)$ & 13.5 & 2.98 & $2920_{-1420}^{+1386}$ & 0.97 & \dots & \dots \\
160 & G288.6551-00.0986 & 163.61027 & -59.68738 & $20.928(11)$ & $18.352(7)$ & $17.004(6)$ & $16.006(5)$ & $15.544(6)$ & $13.680(31)$ & $12.589(17)$ & $12.017(22)$ & $10.139(44)$ & $10.146(57)$ & $10.025(58)$ & $10.029(54)$ & -2.70 & Class III & \dots & \dots & \dots & \dots & \dots & \dots & \dots & 1.09 & \dots & \dots \\
161 & G288.9520-00.6470 & 163.66919 & -60.31033 & $15.686(5)$ & $15.020(7)$ & $14.780(5)$ & $14.647(5)$ & $17.565(21)$ & $14.010(15)$ & $13.105(25)$ & $12.717(22)$ & $11.847(42)$ & $11.828(71)$ & $11.919(89)$ & $11.514(76)$ & -2.51 & Class III & BSC & $7000(50)$ & $4.50(25)$ & $331.118(122)$ & 143.1 & 3.38 & $7306_{-1318}^{+916}$ & 7.69 & 416 & 0.77 \\
162 & G285.5177-00.0328 & 158.30562 & -58.15277 & $21.469(32)$ & $19.747(18)$ & $19.842(20)$ & $18.972(13)$ & $18.850(26)$ & $15.944(20)$ & $14.839(29)$ & $14.231(23)$ & $13.006(61)$ & $12.453(103)$ & $12.106(108)$ & $11.150(79)$ & -0.76 & Class II & BSC & $4200(50)$ & $3.00(25)$ & $0.677(6)$ & 53.6 & 1.66 & $3162_{-652}^{+1284}$ & 1.31 & \dots & \dots \\
163 & G288.2967-00.9833 & 162.18250 & -60.32070 & $22.097(62)$ & $19.738(18)$ & $18.361(11)$ & $17.340(7)$ & $16.881(7)$ & $14.943(19)$ & $13.789(18)$ & $13.286(23)$ & $11.505(87)$ & $11.610(68)$ & $11.554(101)$ & $11.310(101)$ & -2.61 & Class III & \dots & \dots & \dots & \dots & \dots & \dots & \dots & 1.10 & 263 & 0.16 \\
164 & G284.9066-00.0851 & 157.26354 & -57.88515 & $17.631(6)$ & $16.668(5)$ & $16.325(5)$ & $16.090(5)$ & $15.928(5)$ & $14.782(9)$ & $14.057(7)$ & $13.569(24)$ & $12.537(72)$ & $12.285(102)$ & $12.102(102)$ & $11.455(54)$ & -1.63 & Class III & \dots & \dots & \dots & \dots & \dots & \dots & \dots & 1.34 & 517 & 1.53 \\
165 & G284.6047-00.2390 & 156.62817 & -57.85870 & $21.240(14)$ & $18.602(7)$ & $17.250(8)$ & $16.274(7)$ & $15.850(8)$ & $13.995(14)$ & $12.888(23)$ & $12.407(25)$ & $11.285(71)$ & $11.213(76)$ & $11.161(81)$ & $11.260(77)$ & -2.82 & Class III & BS & $3400(50)$ & $4.00(25)$ & $16.624(76)$ & 19.7 & 3.11 & $4429_{-665}^{+625}$ & 0.95 & 511 & 0.59 \\
\hline
\end{tabular}
\end{sidewaystable*}

\addtocounter{table}{-1}
\begin{sidewaystable*}
\caption{Photometric, astrometric, and physical properties of the MIR-excess candidates (continued).}
\centering
\tiny
\setlength{\tabcolsep}{0.8pt}
\renewcommand{\arraystretch}{0.92}
\begin{tabular}{c l c c c c c c c c c c c c c c c c c c c c c c c c c c}
\hline\hline
\# & GLIMPSE ID & RA & Dec & $g$ & $r$ & $i$ & $z$ & $Y$ & $J$ & $H$ & $K_s$ & $[3.6]$ & $[4.5]$ & $[5.8]$ & $[8.0]$ & $\alpha_{\rm IRAC}$ & Class & Model & $T_{\rm eff}$ & $\log g$ & $L_{\rm bol}$ & $V_{\rm gfb}$ & $A_V$ & $D_{\rm photogeo}$ & RUWE & $n_{\rm pairs}$ & $J_{\rm W1W2}$ \\
 &  & (deg) & (deg) & (mag) & (mag) & (mag) & (mag) & (mag) & (mag) & (mag) & (mag) & (mag) & (mag) & (mag) & (mag) &  &  &  & (K) & (dex) & ($L_\odot$) &  & (mag) & (pc) &  &  &  \\
\hline
166 & G286.0209-00.5670 & 158.62436 & -58.86764 & $19.756(8)$ & $17.777(6)$ & $16.854(5)$ & $16.159(5)$ & $15.824(6)$ & $14.231(23)$ & $13.292(26)$ & $13.004(25)$ & $11.520(38)$ & $11.561(61)$ & $11.496(51)$ & $11.512(68)$ & -2.82 & Class III & \dots & \dots & \dots & \dots & \dots & \dots & \dots & 1.04 & 499 & 0.46 \\
167 & G288.2845-00.9811 & 162.16194 & -60.31306 & $18.518(4)$ & $16.839(11)$ & $16.033(5)$ & $15.436(5)$ & $15.169(5)$ & $13.661(21)$ & $12.905(31)$ & $12.540(25)$ & $11.495(63)$ & $11.583(69)$ & $11.507(107)$ & $11.221(103)$ & -2.52 & Class III & BSC & $4600(50)$ & $3.00(25)$ & $43.973(119)$ & 16.7 & 2.62 & $5844_{-729}^{+733}$ & 1.02 & 478 & 0.50 \\
168 & G282.3402-00.1902 & 153.17690 & -56.57807 & $16.892(5)$ & $15.740(4)$ & $14.745(6)$ & $15.751(7)$ & $14.447(6)$ & $13.285(28)$ & $12.573(30)$ & $12.369(26)$ & $11.573(49)$ & $11.504(72)$ & $11.479(85)$ & $11.521(83)$ & -2.80 & Class III & BSC & $4500(50)$ & $4.50(25)$ & $0.086(0)$ & 49.4 & 0.13 & $357_{-15}^{+14}$ & 3.03 & 521 & 0.52 \\
169 & G282.5524-00.1661 & 153.51988 & -56.67833 & $23.015(46)$ & $19.839(9)$ & $18.147(7)$ & $16.968(6)$ & $16.484(7)$ & $14.561(10)$ & $13.523(9)$ & $13.166(26)$ & $12.501(65)$ & $12.397(94)$ & $11.975(108)$ & $12.079(104)$ & -2.28 & Class III & BS & $3400(50)$ & $4.00(25)$ & $6.538(13)$ & 1.5 & 4.74 & $2948_{-578}^{+1359}$ & 0.93 & \dots & \dots \\
170 & G288.8191-00.7217 & 163.36162 & -60.31969 & $16.420(6)$ & $15.610(7)$ & $15.237(6)$ & $14.833(6)$ & $14.681(6)$ & $15.725(82)$ & $13.924(26)$ & $13.126(27)$ & $11.890(85)$ & $11.816(61)$ & $11.664(94)$ & $11.707(83)$ & -2.62 & Class III & BS & $10000(100)$ & $3.00(25)$ & $244.790(94)$ & 292.2 & 3.04 & $6098_{-1029}^{+1739}$ & 1.06 & 444 & 0.74 \\
171 & G286.7479-00.9365 & 159.50064 & -59.54821 & $20.871(12)$ & $18.583(7)$ & $17.471(6)$ & $16.605(5)$ & $16.225(7)$ & $14.503(20)$ & $13.527(16)$ & $13.117(28)$ & $11.588(69)$ & $11.503(85)$ & $11.362(88)$ & $11.785(100)$ & -3.04 & Class III & \dots & \dots & \dots & \dots & \dots & \dots & \dots & 0.99 & 483 & 0.45 \\
172 & G287.9435-00.4509 & 162.03764 & -59.68592 & $18.768(6)$ & $16.876(6)$ & $15.897(7)$ & $15.361(6)$ & $15.068(6)$ & $13.570(20)$ & $12.751(24)$ & $12.361(28)$ & $11.556(42)$ & $11.491(65)$ & $11.673(99)$ & $11.537(91)$ & -2.89 & Class III & BSC & $4300(50)$ & $4.50(25)$ & $28.609(62)$ & 8.0 & 2.62 & $4572_{-851}^{+729}$ & 1.50 & 206 & 0.89 \\
173 & G283.2995-00.9496 & 153.84753 & -57.74624 & $18.911(6)$ & $16.830(5)$ & $15.834(6)$ & $15.089(6)$ & $16.694(12)$ & $13.147(24)$ & $12.295(20)$ & $11.972(29)$ & $11.034(88)$ & $10.926(93)$ & $10.844(67)$ & $11.143(93)$ & -2.96 & Class III & BSC & $6300(50)$ & $4.50(25)$ & $29.721(88)$ & 77.7 & 3.47 & $3773_{-463}^{+502}$ & 1.05 & 496 & 0.36 \\
174 & G287.5672-00.2963 & 161.51853 & -59.37570 & $22.240(38)$ & $20.350(24)$ & $18.996(10)$ & $18.113(7)$ & $17.791(9)$ & $16.218(8)$ & $15.401(11)$ & $14.940(33)$ & $13.158(73)$ & $12.358(80)$ & $11.669(93)$ & $10.680(65)$ & -0.01 & Flat & BS & $3400(50)$ & $3.50(25)$ & $1.333(8)$ & 13.6 & 2.23 & $3834_{-814}^{+1263}$ & 1.06 & 11 & 2.32 \\
175 & G284.5883-00.1810 & 156.65941 & -57.80059 & $19.402(6)$ & $17.209(4)$ & $16.176(5)$ & $15.459(5)$ & $15.118(5)$ & $13.533(26)$ & $12.586(39)$ & $12.214(35)$ & $10.170(44)$ & $10.179(39)$ & $10.137(50)$ & $10.100(30)$ & -2.76 & Class III & BSC & $4100(50)$ & $3.00(25)$ & $21.893(83)$ & 28.1 & 3.14 & $3498_{-390}^{+431}$ & 1.07 & 516 & 0.46 \\
176 & G284.8721-00.6628 & 156.63702 & -58.35986 & $22.449(35)$ & $19.657(9)$ & $18.147(7)$ & $17.061(6)$ & $16.611(6)$ & $14.996(44)$ & $13.482(34)$ & $12.688(35)$ & $11.620(46)$ & $11.552(69)$ & $11.511(96)$ & $11.534(83)$ & -2.75 & Class III & BSC & $4800(50)$ & $3.50(25)$ & $114.502(156)$ & 14.9 & 6.72 & $7732_{-3148}^{+1943}$ & 1.43 & 525 & 0.45 \\
177 & G283.1939-00.5288 & 154.12150 & -57.33825 & $19.314(8)$ & $17.600(5)$ & $16.668(6)$ & $15.966(6)$ & $15.655(6)$ & $14.252(48)$ & $13.546(52)$ & $13.380(36)$ & $10.948(41)$ & $10.973(63)$ & $10.949(64)$ & $11.050(80)$ & -2.96 & Class III & \dots & \dots & \dots & \dots & \dots & \dots & \dots & 1.07 & 494 & 0.64 \\
178 & G284.8724-00.6696 & 156.63048 & -58.36586 & $23.571(82)$ & $20.279(11)$ & $18.568(8)$ & $17.287(6)$ & $16.743(6)$ & $14.643(19)$ & $13.329(18)$ & $12.748(36)$ & $11.676(46)$ & $11.514(72)$ & $11.498(84)$ & $11.440(92)$ & -2.60 & Class III & BS & $2600(50)$ & $3.00(25)$ & $0.084(1)$ & 95.2 & 0.31 & $607_{-97}^{+4214}$ & 1.61 & 526 & 0.51 \\
179 & G282.4611-00.3689 & 153.17327 & -56.79377 & $22.990(61)$ & $19.705(11)$ & $17.912(8)$ & $16.679(7)$ & $16.152(8)$ & $14.107(6)$ & $12.872(9)$ & $12.803(37)$ & $11.959(72)$ & $11.793(65)$ & $11.794(108)$ & $11.684(82)$ & -2.57 & Class III & BS & $3600(50)$ & $3.50(25)$ & $12.353(19)$ & 5.1 & 5.57 & $2895_{-602}^{+1631}$ & 0.99 & 515 & 0.19 \\
180 & G284.1733-00.5057 & 155.67457 & -57.85527 & $22.511(56)$ & $20.536(24)$ & $19.487(20)$ & $18.740(14)$ & $18.393(18)$ & $16.935(11)$ & $16.212(17)$ & $15.872(42)$ & $13.383(84)$ & $12.768(79)$ & $10.655(61)$ & $8.819(45)$ & 2.70 & Class I & BSC & $5500(50)$ & $3.00(25)$ & $4.781(15)$ & 12.8 & 4.49 & $5605_{-2191}^{+1586}$ & 1.08 & 75 & 0.14 \\
181 & G283.3939-00.4398 & 154.52189 & -57.37501 & $16.724(5)$ & $15.854(5)$ & $15.607(6)$ & $15.460(6)$ & $15.361(6)$ & $14.893(31)$ & $13.924(28)$ & $13.824(42)$ & $11.498(51)$ & $10.926(56)$ & $10.627(53)$ & $10.080(48)$ & -1.26 & Class II & BSC & $7000(50)$ & $4.50(25)$ & $7.797(12)$ & 19.6 & 1.73 & $2527_{-458}^{+515}$ & 4.51 & 490 & 0.60 \\
182 & G288.3481-00.1058 & 163.05910 & -59.55916 & $20.418(11)$ & $17.808(8)$ & $16.527(5)$ & $15.633(5)$ & $15.229(5)$ & $13.982(32)$ & $13.253(39)$ & $12.857(42)$ & $11.243(39)$ & $11.278(63)$ & $11.035(74)$ & $11.188(55)$ & -2.72 & Class III & BS & $3400(50)$ & $4.50(25)$ & $23.539(76)$ & 17.1 & 2.77 & $4838_{-946}^{+890}$ & 1.70 & 472 & 0.64 \\
183 & G284.9481-00.9076 & 156.51379 & -58.60799 & $13.923(10)$ & $13.825(10)$ & $13.733(10)$ & $13.869(10)$ & $13.748(6)$ & $13.703(45)$ & $12.917(44)$ & $12.599(46)$ & $11.818(51)$ & $11.737(63)$ & $11.597(72)$ & $11.537(78)$ & -2.51 & Class III & BS & $12000(175)$ & $4.50(25)$ & $362.058(80)$ & 38.8 & 2.55 & $3482_{-483}^{+562}$ & 6.56 & 348 & 1.26 \\
184 & G285.2308-00.3977 & 157.48317 & -58.32023 & $21.101(11)$ & $18.594(5)$ & $17.291(5)$ & $16.338(5)$ & $15.912(5)$ & $15.714(63)$ & $14.051(58)$ & $13.436(46)$ & $11.862(29)$ & $11.825(82)$ & $11.613(63)$ & $11.336(82)$ & -2.22 & Class III & BS & $3800(50)$ & $4.50(25)$ & $11.188(32)$ & 90.7 & 3.64 & $3979_{-1319}^{+1213}$ & 1.07 & 499 & 0.59 \\
185 & G286.0974-00.0889 & 159.20882 & -58.49069 & $22.239(24)$ & $19.940(9)$ & $18.292(6)$ & $17.365(5)$ & $16.856(6)$ & $15.027(20)$ & $13.842(24)$ & $13.303(49)$ & $11.950(54)$ & $11.436(61)$ & $10.902(80)$ & $10.045(49)$ & -0.66 & Class II & BS & $3500(50)$ & $3.00(25)$ & $3.545(17)$ & 16.2 & 3.16 & $3299_{-799}^{+1383}$ & 2.89 & 478 & 1.43 \\
186 & G288.3480-00.8904 & 162.35909 & -60.26109 & $22.975(53)$ & $20.991(19)$ & $20.510(24)$ & $19.681(15)$ & $19.410(26)$ & $19.602(84)$ & $17.600(45)$ & $16.245(53)$ & $12.424(54)$ & $11.745(74)$ & $11.142(76)$ & $10.681(47)$ & -0.84 & Class II & BSC & $7000(50)$ & $3.00(25)$ & $14.625(120)$ & 122.0 & 4.46 & $12626_{-2706}^{+1818}$ & 1.15 & 119 & 1.03 \\
187 & G284.5821-00.7275 & 156.10307 & -58.26118 & $18.269(6)$ & $16.398(5)$ & $15.534(5)$ & $14.909(5)$ & $14.600(5)$ & $15.341(58)$ & $14.043(47)$ & $13.492(59)$ & $11.489(41)$ & $11.517(64)$ & $11.571(106)$ & $11.537(90)$ & -2.92 & Class III & BSC & $4900(50)$ & $4.50(25)$ & $33.641(50)$ & 300.4 & 2.78 & $4481_{-375}^{+444}$ & 0.99 & 532 & 0.63 \\
188 & G286.0633-00.5608 & 158.70145 & -58.88321 & $19.229(7)$ & $17.011(6)$ & $15.939(6)$ & $15.134(6)$ & $14.747(7)$ & $15.582(101)$ & $14.214(60)$ & $13.725(74)$ & $11.376(47)$ & $11.429(65)$ & $11.088(69)$ & $11.222(71)$ & -2.58 & Class III & BSC & $6200(50)$ & $4.50(25)$ & $255.344(167)$ & 453.4 & 4.85 & $7691_{-2350}^{+1972}$ & 1.34 & 505 & 0.56 \\
189 & G283.8449-00.9989 & 154.64924 & -58.09047 & $21.391(15)$ & $19.539(8)$ & $18.696(8)$ & $18.006(7)$ & $17.809(10)$ & $16.256(31)$ & $15.489(55)$ & $15.027(75)$ & $11.577(46)$ & $11.503(78)$ & $11.083(84)$ & $11.239(86)$ & -2.37 & Class III & BSC & $4100(50)$ & $3.00(25)$ & $1.156(16)$ & 41.4 & 2.26 & $2574_{-551}^{+885}$ & 1.23 & 499 & 0.57 \\
190 & G288.0846-00.1563 & 162.54936 & -59.48690 & $20.604(9)$ & $17.899(6)$ & $16.517(6)$ & $15.482(5)$ & $15.033(5)$ & $14.565(49)$ & $13.792(84)$ & $13.661(79)$ & $11.029(37)$ & $11.052(66)$ & $10.966(72)$ & $10.892(73)$ & -2.68 & Class III & BSC & $4600(50)$ & $4.50(25)$ & $89.711(119)$ & 278.1 & 5.14 & $5578_{-2002}^{+1503}$ & 1.33 & 474 & 0.64 \\
191 & G283.8879-00.5994 & 155.12911 & -57.77960 & $18.927(8)$ & $17.058(5)$ & $16.173(7)$ & $15.239(7)$ & $14.838(7)$ & $14.396(59)$ & $13.628(64)$ & $13.274(93)$ & $11.100(41)$ & $10.958(45)$ & $10.943(79)$ & $10.608(66)$ & -2.32 & Class III & BS & $3200(50)$ & $4.50(25)$ & $0.410(3)$ & 41.1 & 0.36 & $906_{-148}^{+255}$ & 6.61 & 484 & 10.08 \\
\hline
\end{tabular}
\vspace{1mm}
\begin{minipage}{0.98\textwidth}
\tiny
\raggedright
\setlength{\parindent}{0pt}
\textbf{Notes.} (1) Source number in this table. (2) GLIMPSE identifier. The common prefix ``SSTVELA'' has been omitted for brevity; all sources belong to the SSTVELA catalog. (3--4) Right ascension and declination in the ICRS (J2000) reference frame. (5--9) DECaPS $g$, $r$, $i$, $z$, and $Y$ magnitudes. (10--12) VVVX $J$, $H$, and $K_s$ magnitudes. (13--16) \textit{Spitzer}/IRAC magnitudes at 3.6, 4.5, 5.8, and 8.0~$\mu$m. For quantities with symmetric uncertainties, the notation $x(y)$ is adopted, where $y$ represents the $1\sigma$ uncertainty in the last quoted digit or digits of $x$; for example, $15.879(5)$ corresponds to $15.879\pm0.005$. (17) IRAC spectral index $\alpha_{\rm IRAC}$, derived from the four IRAC bands. (18) IRAC spectral-index class following \citet{Greene1994}. (19) Atmospheric model used in the VOSA fitting: BS = BT-Settl \citep{Allard2012,Asplund2009}; BSC = BT-Settl CIFIST \citep{Allard2012,Caffau2011}; and KU03 = Kurucz ATLAS9 ODFNEW/NOVER \citep{CastelliKurucz2003}. (20) Effective temperature $T_{\rm eff}$ and its uncertainty from the VOSA SED fitting. (21) Surface gravity $\log g$ and its uncertainty from the VOSA SED fitting. (22) Bolometric luminosity $L_{\rm bol}$ and its uncertainty from the VOSA SED fitting. (23) Modified goodness-of-fit statistic $V_{\rm gfb}$ provided by VOSA. (24) Line-of-sight visual extinction $A_V$ from the three-dimensional dust map of Z25. The $A_V$ values are the Z25 line-of-sight extinction estimates evaluated at the median BJ21 photogeometric distances. No independent formal uncertainty on $A_V$ was available from the adopted extraction, and therefore no artificial uncertainty is assigned. (25) BJ21 photogeometric posterior median distance $D_{\rm photogeo}$; the lower and upper uncertainties are derived from the 16th and 84th percentiles, respectively. For sources without a valid BJ21 photogeometric distance estimate, the VOSA-derived physical parameters (model atmosphere, $T_{\rm eff}$, $\log g$, $L_{\rm bol}$, and $V_{\rm gfb}$), the Z25 extinction, and the photogeometric distance are omitted because the SED fitting relies on physically meaningful distance and extinction estimates. (26) Gaia renormalised unit weight error (RUWE), included as an astrometric-quality diagnostic. (27) Number of paired NEOWISE $W1$ and $W2$ observations used to calculate the variability index. (28) Stetson variability index $J_{\rm W1W2}$ computed from contemporaneous NEOWISE $W1$ and $W2$ measurements. The complete machine-readable table, including auxiliary variability and cross-identification parameters, is available at the CDS.
\end{minipage}
\end{sidewaystable*}

\FloatBarrier

% --- Declaration of Interest ---
\clearpage
\section*{Declaration of Interest}

The authors declare that they have no known competing financial interests or personal relationships that could have appeared to influence the work reported in this paper.

% --- References ---

\bibliographystyle{elsarticle-harv}
\bibliography{references}

\end{document}